\documentclass[twocolumn,floatfix,tightenlines,superscriptaddress]{revtex4}
\usepackage{mathtools}
\usepackage{bm}
\usepackage{dsfont,amsthm,amsbsy}
\usepackage{verbatim}
\usepackage{amssymb}
\usepackage{amsmath}
\usepackage{bbm}
\usepackage{graphicx}
\usepackage{epstopdf}
\usepackage{subfigure}
\usepackage{natbib}
\usepackage{epsfig}
\usepackage{amsfonts}
\usepackage{mathrsfs}
\usepackage{sidecap}
\usepackage{lipsum}
\usepackage[toc,page,title,titletoc,header]{appendix}
\usepackage[colorlinks,linkcolor=blue,citecolor=blue,anchorcolor=blue,urlcolor=blue]{hyperref}
\usepackage{hyperref}
\usepackage{resizegather}
\usepackage{tikz}
\usepackage{float}
\usepackage{mathbbol}

\usepackage[normalem]{ulem}
\usepackage{cancel}
\usepackage{upgreek}

\usepackage{dblfloatfix}
\usepackage{graphicx}  
\usepackage{dcolumn}   
\usepackage{bm}        
\usepackage{amssymb}   
\usepackage{amsmath}
\usepackage{lipsum}
\usepackage{xcolor}
\usepackage{ulem}
\usepackage{braket}

\begin{document}
	
	\title{Emergence of prethermal time quasicrystalline order in a quasiperiodically driven non-interacting spin chain}
	 \author{Davood Marripour}
      \affiliation{Department of Physics, Institute for Advanced Studies in Basics Sciences (IASBS), Zanjan 45137-66731, Iran}
     \author{Jahanfar Abouie}
     \email{jahan@iasbs.ac.ir}
    \affiliation{Department of Physics, Institute for Advanced Studies in Basics Sciences (IASBS), Zanjan 45137-66731, Iran}

    \date{\today}

	\begin{abstract}
	We study prethermal time quasicrystalline (TQC) order in a quasiperiodically driven chain of non-interacting spin-1/2 particles. The drive consists of two parts, switched on and off periodically with frequency $\omega_d$: (i) disordered Ising interactions, with exchange couplings chosen from a symmetric interval $[-J/2, J/2]$, allowing random antiferromagnetic or ferromagnetic nearest-neighbor couplings, together with a random transverse field; and (ii) a rotating transverse magnetic field with frequency $\Omega$.
	The ratio $\omega_d/\Omega$ is chosen to be irrational, producing multiple incommensurate frequencies and yielding quasiperiodic dynamics beyond Floquet theory. Using exact diagonalization, we analyze the time autocorrelation function, dynamical structure factor, and entanglement entropy (EE). In the high-frequency regime, robust spectral peaks at incommensurate frequencies (not integer multiples of the fundamental drives) signal quasiperiodic time-translation symmetry breaking (QTTSB). The EE exhibits sublinear power-law growth followed by a prethermal plateau, indicating suppressed resonant heating due to an energy scale mismatch. The nonequilibrium lifetime increases rapidly with driving frequency. Unlike symmetric disorder sampling, an asymmetric distribution of the Ising exchange couplings induces collective spin rigidity, enhancing the system's resistance to heating. The TQC phase remains stable against next-nearest-neighbor (NNN) exchange perturbations and rotational imperfections, with robustness comparable to discrete time crystals (TCs) under periodic driving. Our results establish this quasiperiodically driven system as a platform for long-lived nonequilibrium temporal order, revealing the interplay of disorder, collective rigidity, and quasiperiodic driving.
\end{abstract}

	\maketitle
	
	\section{Introduction}
The nonequilibrium behavior of quantum systems under time-dependent driving has attracted widespread interest in recent years \cite{r1, r2, r3, r4, r5, r6, r7, r8, r9, r10, r11, r12, r13, r14, r15, r16, r17,px63-dtc9}. This context gives rise to a variety of remarkable phenomena, including time crystals (TCs) \cite{r18,r19,r20,r21,r22,r23,r24,r25,r26,r27,marripour,Y. Huang,Khemani2019,r29,r30}, time quasicrystals (TQCs) \cite{r31,r32,r33,r34,r35,r36,r37,r38,r39,r40,r41,r42,r43,r44}, dynamical freezing \cite{das2010,bhattacharyya2012,hegde2014,mondal2012,guo2025,divakaran2014,camilo2020,mukherjee2024,koch2023,haldar2021,banerjee2023a,banerjee2024,turkeshi2023,gangopadhay2025}, quantum many-body scars \cite{pai2019,mukherjee2020a,mizuta2020,H. Yarloo,sugiura2021,mukherjee2020b,maskara2021,hudomal2022,huang2022}, Floquet prethermalization \cite{B. Bauer,K. Mallayya,P. T. Dumitrescu,D. V. Else,G. He,S. A. Weidinger,E. Canovi,M. Bukov,Kyprianidis,Das Sarma,Zeng,Stasiuk}, and many-body localization (MBL) \cite{Johri,Keyserlingk,Oganesyan,Huse,Nandkishore,Lazarides,Kjall,huse,Bordia,P. Ponte}.

Understanding the nonequilibrium dynamics of quantum many-body systems is a central challenge in modern condensed matter physics \cite{Igloi2007}. Mitigating heating is essential to prevent the system from reaching an infinite temperature ensemble, where temporal correlations vanish. Proposed mechanisms to suppress heating and break the eigenstate thermalization hypothesis (ETH) include MBL \cite{Johri,Huse,Nandkishore,huse}, prethermalization at high driving frequencies \cite{B. Bauer,K. Mallayya,D. V. Else,M. Bukov}, quantum many-body scars \cite{pai2019,mukherjee2020a,mizuta2020,H. Yarloo,sugiura2021,mukherjee2020b,maskara2021,hudomal2022,huang2022}, Hilbert space fragmentation \cite{Langlett,Moudgalya,Moudgalya1}, and electric-field-driven dynamical localization \cite{Y. Baum,Luitz,Agarwala,T. Nag,L. Tamang,M. Fava,Keser,Martinez}.

When heating is successfully suppressed, driven quantum systems can host robust nonequilibrium phases with no equilibrium analogs. Among the most fascinating are TCs and TQCs, emerging under periodic and quasiperiodic driving, respectively. While TCs break discrete time-translation symmetry via subharmonic response, TQCs exhibit a richer temporal order: they respond at multiple incommensurate frequencies that are not integer multiples of the driving frequency, resulting in quasiperiodic dynamics akin to spatial quasicrystals \cite{r27,marripour,r38}. This nonequilibrium phase is fundamentally characterized by QTTSB. In frequency space, this symmetry breaking is directly observable in the Fourier spectrum of local observables: the system's response develops sharp, rigid peaks at new incommensurate frequencies, generating a distinct quasiperiodic temporal lattice that differs from the driving protocol \cite{r27,marripour}.

TQCs arise from quasiperiodic driving, where the time-dependent Hamiltonian $H(t)$ incorporates incommensurate frequencies. This yields nonrepetitive evolution that densely explores a multidimensional toroidal phase space without returning to a previous configuration, enabling intricate dynamics such as fractal quasienergy spectra, Poissonian level statistics (indicative of nonergodic localized dynamics), and dense phase-space trajectories \cite{D. V. Else,P. T. Dumitrescu}.

However, realizing TQCs in many-body systems faces a fundamental stability challenge. Like periodically driven systems, which generically heat to a featureless infinite-temperature state \cite{Lazarides2014,DAlessio2014}, quasiperiodically driven systems are even more prone to thermalization due to their dense, multi-frequency spectrum providing additional resonance channels. Stabilizing TQCs thus requires robust ergodicity-breaking mechanisms, such as MBL, to prevent thermalization from washing out coherent quantum dynamics.
  
Despite this, quasiperiodic driving offers a promising route to nonequilibrium phases beyond the Floquet paradigm, as it can strongly modify energy absorption and dynamical correlations \cite{Martin2021,V. Tiwari}. While quasiperiodic driving ultimately leads to thermalization, this occurs only after a long prethermal time, analogous to high-frequency periodic driving where the system relaxes into a long-lived metastable state before eventual thermalization \cite{P. T. Dumitrescu,Zhao2021,D. V. Else,G. He,S. A. Weidinger,E. Canovi,M. Bukov}.
	
In disordered spin chains subjected to a quasiperiodic drive (e.g., via a stroboscopic Fibonacci sequence), studies reveal that after an initial transient, the system enters a long-lived glassy regime characterized by logarithmic EE growth and extremely slow decay of spin correlations—behavior reminiscent of the MBL transition \cite{P. T. Dumitrescu,r35}. Although the system ultimately heats to infinite temperature on timescales that grow exponentially for weak or high-frequency drives, this slow relaxation enables metastable dynamical phases, including TQCs, where spins exhibit persistent quasiperiodic oscillations distinct from the drive \cite{P. T. Dumitrescu}. Similar TQC behavior has been demonstrated in a disordered quantum Ising chain under a quasiperiodic transverse field, where transverse magnetization shows quasiperiodic oscillations persisting over extended prethermal timescales, indicating that the TQC exists as a long-lived prethermal dynamical phase rather than a true equilibrium state \cite{marripour}.

In this paper, we study the emergence of prethermal TQC order in a chain of non-interacting spin-1/2 particles subjected to quasiperiodic driving. The drive consists of two components, switched on and off periodically with frequency $\omega_d$: (i) disordered Ising interactions—with exchange couplings randomly drawn from a symmetric interval $[-J/2, J/2]$, allowing for both antiferromagnetic and ferromagnetic nearest-neighbor couplings—together with a random transverse field; and (ii) a rotating transverse magnetic field with frequency $\Omega$. The ratio $\omega_d/\Omega$ is chosen to be irrational, generating multiple incommensurate frequencies and thereby producing quasiperiodic dynamics that extend beyond the conventional Floquet framework.

We analyze the system's dynamical observables—including time autocorrelation functions, local magnetization, and the growth of EE—across different frequency regimes to identify the conditions under which a long-lived prethermal TQC phase emerges. EE serves as a particularly useful diagnostic to distinguish dynamical phases. In the MBL phase, EE grows logarithmically as $S(t) \sim \log(t)$. In thermalizing systems, it grows linearly as $S(t) \sim t$, signaling rapid entanglement spreading and eventual thermal equilibrium. Under quasiperiodic driving in the prethermal regime, however, EE exhibits sublinear power-law growth $S(t) \sim t^{\alpha}$ with $\alpha<1$, indicating a long-lived non-thermal quasistationary state characterized by complex quasiperiodic oscillations that prevent full thermalization \cite{P. T. Dumitrescu, Zhao2021, M. Serbyn, X. Wen}.

Furthermore, we demonstrate how the time autocorrelation function and the EE behavior can distinguish the TQC phase from the system's eventual thermalization to infinite temperature. Our analysis reveals that the lifetime of the prethermal TQC regime increases rapidly with driving frequency, consistent with the suppression of resonant energy absorption. We also compare the effects of symmetric versus asymmetric disorder sampling, showing that asymmetric distributions induce collective spin rigidity, which enhances the system's resistance to heating.

The remainder of this paper is organized as follows. In Sec.~\ref{Sec:our model}, we introduce the model Hamiltonian and describe the quasiperiodic driving protocol. In this section, we present our numerical results for the time autocorrelation function and dynamical structure factor, demonstrating QTTSB. In Sec.~\ref{Sec:entanglement entropy}, we analyze the EE dynamics and characterize the prethermal plateau and its frequency dependence. In Sec.~\ref{Sec:Rigidity}, we examine the stability of the TQC phase against next-nearest-neighbor perturbations and rotational imperfections. Finally, in Sec.~\ref{Sec:Conclution}, we summarize our findings and discuss open questions.

\section{Quasiperiodically driven spin chain} \label{Sec:our model}

We consider a non-interacting spin-1/2 chain subjected to a quasiperiodic drive. The drive consists of two parts, switched on and off periodically with frequency $\omega_d$: (i) disordered Ising interactions with exchange couplings drawn from a symmetric interval $[-J/2, J/2]$ (random antiferromagnetic or ferromagnetic nearest-neighbor couplings) together with a random transverse field; and (ii) a transverse magnetic field oscillating with frequency $\Omega$ in the $xy$-plane. The ratio $\omega_d/\Omega$ is chosen to be irrational, producing multiple incommensurate frequencies and yielding quasiperiodic dynamics.

Within this framework, the time-dependent Hamiltonian is given by:
\begin{equation}
    \hat{H}(t) = \hat{H}_{\text{ITF}}(t) + \hat{H}_{\text{kick}}(t) + \frac{\Omega}{2} \sum_i \sigma^z_i,
    \label{eq:H_total}
\end{equation}
where the first term is the time-modulated disordered transverse-field Ising (ITF) Hamiltonian:
\begin{equation}
    \hat{H}_{\text{ITF}}(t) = h_1(t) \sum_{i=1}^{L} \left( J_i \sigma^z_i \sigma^z_{i+1} + h_i \sigma^x_i \right),
    \label{eq:H_ITF}
\end{equation}
with $\sigma_i^{\alpha}$ ($\alpha = x, z$) Pauli operators, $J_i \in [-J/2, J/2]$ the random nearest-neighbor coupling, and $h_i \in [0, h]$ a random transverse field. The second term is the kick Hamiltonian:
\begin{equation}
    \hat{H}_{\text{kick}}(t) = h_2(t) \sum_i \left( \cos(\Omega t)\,\sigma^y_i - \sin(\Omega t)\,\sigma^x_i \right),
    \label{eq:H_drive}
\end{equation}
which rotates the spins in the $xy$-plane.

The time-dependent coefficients $h_1(t)$ and $h_2(t)$ alternate periodically:
\begin{equation}
    \begin{aligned}
        h_1(t) &= 
        \begin{cases}
            0, & \text{for } \frac{T}{4} < t < -\frac{T}{4} \\
            1, & \text{for } -\frac{T}{4} \le t \le \frac{T}{4}
        \end{cases}, \\
        h_2(t) &= 
        \begin{cases}
            \frac{\pi}{T}, & \text{for } \frac{T}{4} < t < -\frac{T}{4} \\
            0, & \text{for } -\frac{T}{4} \le t \le \frac{T}{4}
        \end{cases},
    \end{aligned}
    \label{Eq:h1-h2}
\end{equation}
defining a period $T$ and drive frequency $\omega_d = 2\pi/T$. Quasiperiodic driving is achieved when the ratio $\Omega/\omega_d$ is irrational. The terms $h_2(t) \cos(\Omega t)$ and $h_2(t) \sin(\Omega t)$ in Eq.~\eqref{eq:H_drive} satisfy the quasiperiodicity condition (see Eq. (\ref{Eq:quasiperiod-Hamiltonian}) in Appendix \ref{sec:app}), as evident from their Fourier expansion:
\begin{equation}
    h_2(t)\cos(\Omega t) = \sum_{n_1=-\infty}^{\infty} \sum_{n_2=\pm 1} c_{n_1} e^{i (n_1\omega_d + n_2\Omega) t},
    \label{Eq:quasi-Hamiltonian-drive-expansion}
\end{equation}
where $c_{n_1}$ are Fourier coefficients. These coefficients determine the detailed structure of the quasiperiodic drive, while the interplay of the combined frequencies $n_1\omega_d+n_2\Omega$ gives rise to rich and potentially complex dynamical behavior \cite{r38, marripour}.

In quasiperiodically driven systems, the absence of discrete time-translation symmetry precludes the standard stroboscopic simplifications used in Floquet systems. Consequently, the dynamics must be analyzed via real-time evolution.

The probability distribution of the exchange couplings in the disordered ITF model fundamentally dictates the system's global phase and dynamics. For non-symmetric random interactions, e.g., drawing $J_i$ from $[J/2, 3J/2]$ with $J>0$, the system remains in a globally ordered ferromagnetic phase. Previous studies have shown that under quasiperiodic driving, such systems can host a long-lived prethermal TQC phase \cite{marripour}.

In contrast, drawing $J_i$ from a symmetric distribution $J_i \in [-J/2, J/2]$ introduces competition between random ferromagnetic and antiferromagnetic bonds. This model is gauge-equivalent to a random ferromagnet (a Mattis glass) and maps to non-interacting free fermions, exhibiting single-particle Anderson localization.

Building on this inherently localized structure, we show that under quasiperiodic driving the system avoids rapid thermalization and remains trapped in a long-lived non-equilibrium regime, whose lifetime is strongly controlled by the driving frequency. The persistence of this state—characterized by stable subharmonic temporal oscillations over extended time scales—provides a clear dynamical signature of a prethermal TQC.
	\begin{figure}[htb]
		\centering
		\includegraphics[scale=0.3]{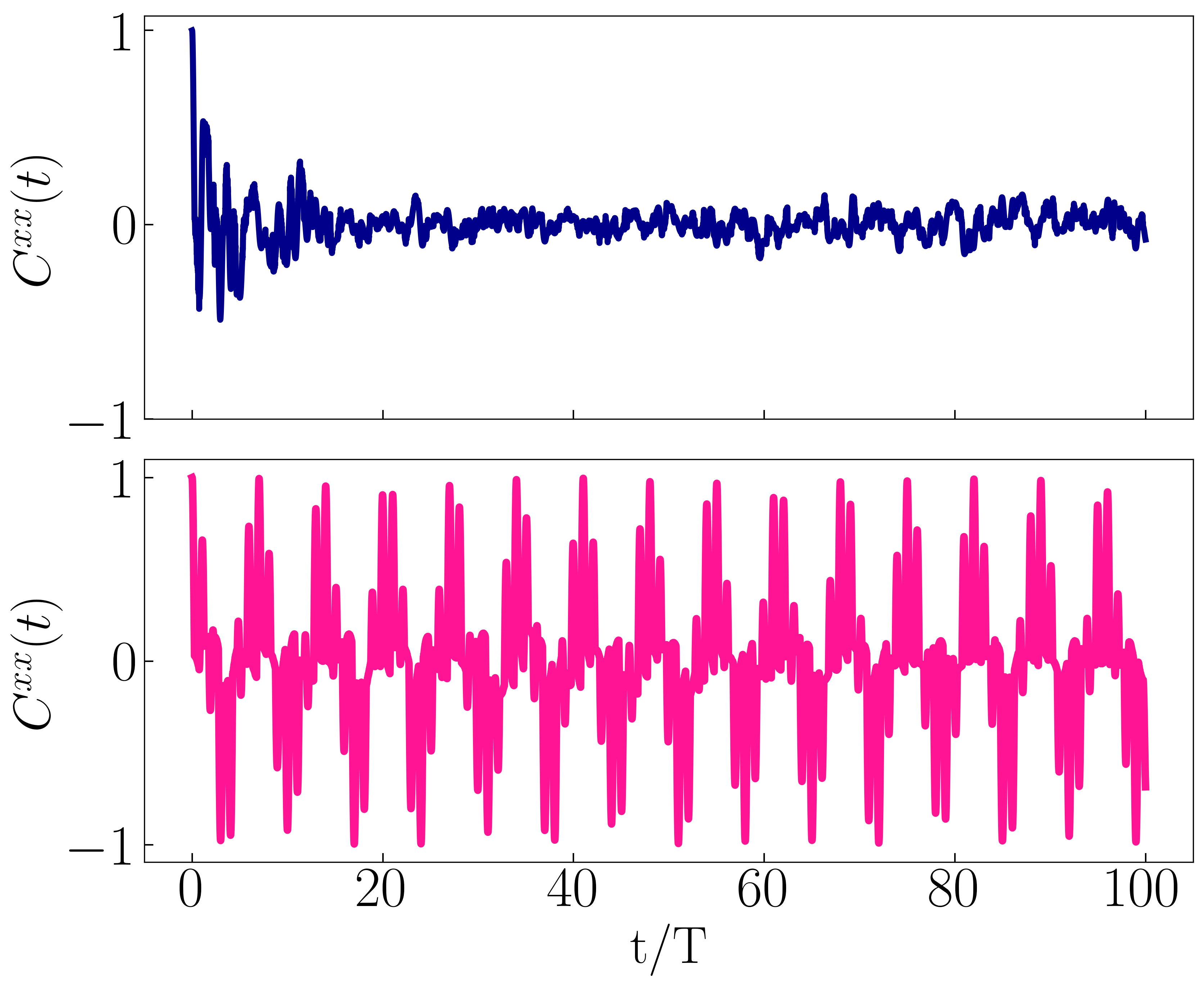}
		\caption{Time autocorrelation function $C^{xx}(t)$ for a single chain of length $L = 6$ with $J=5.5$ and $h=0.3$ under different driving frequencies. \textbf{Top panel} ($\omega_d = 1$): Rapid decay of correlations indicates the absence of long-lived order, reflecting fast energy absorption from the drive at low frequencies. \textbf{Bottom panel} ($\omega_d = 12$): Quasiperiodic oscillations persist over long times, demonstrating the stabilization of non-equilibrium dynamics due to suppressed energy absorption at high driving frequencies.}
		\label{fig:C_diff_omega}
	\end{figure}
	
Before discussing the specific signatures of QTTSB, it is important to establish the driving frequency protocol used in our model. Throughout this analysis, we investigate various frequency regimes, ranging from low to high driving frequencies, while strictly keeping the ratio of the two incommensurate frequencies fixed at $\Omega/\omega_d = \sqrt{2}/4$. We emphasize that this specific choice of an irrational ratio is made for concreteness and does not compromise the generality of the underlying quasiperiodic physics.
	
\textbf{QTTSB}: The phenomenon of QTTSB can be characterized through the temporal autocorrelation function, which serves as a sensitive probe of dynamical stability and memory in disordered driven systems. It is defined as
	\begin{equation}
		C^{\alpha\alpha}(t) = \frac{1}{L} \sum_{i=1}^{L} \overline{\big\langle \hat{\sigma}_i^{\alpha}(t)\,\hat{\sigma}_i^{\alpha}(0) \big\rangle},
		\label{eq:autocorrelation}
	\end{equation}
	where $\alpha\in\{x,y,z\}$, $\langle \cdots \rangle$ denotes the quantum expectation value with respect to the chosen initial state, and the overbar indicates an average over different random realizations.
	
	By varying the overall scale of the frequencies while maintaining their fixed irrational ratio, we can observe distinct dynamical regimes. For instance, as shown in Fig.~\ref{fig:C_diff_omega}, for the low-frequency regime ($\omega_d=1$), the autocorrelation function rapidly decays to zero, indicating a loss of temporal memory. This decay reflects the nature of the external drive at large time periods, where the system absorbs energy efficiently from the drive and is consequently driven toward thermalization.

\begin{figure}
	\includegraphics[scale=0.2]{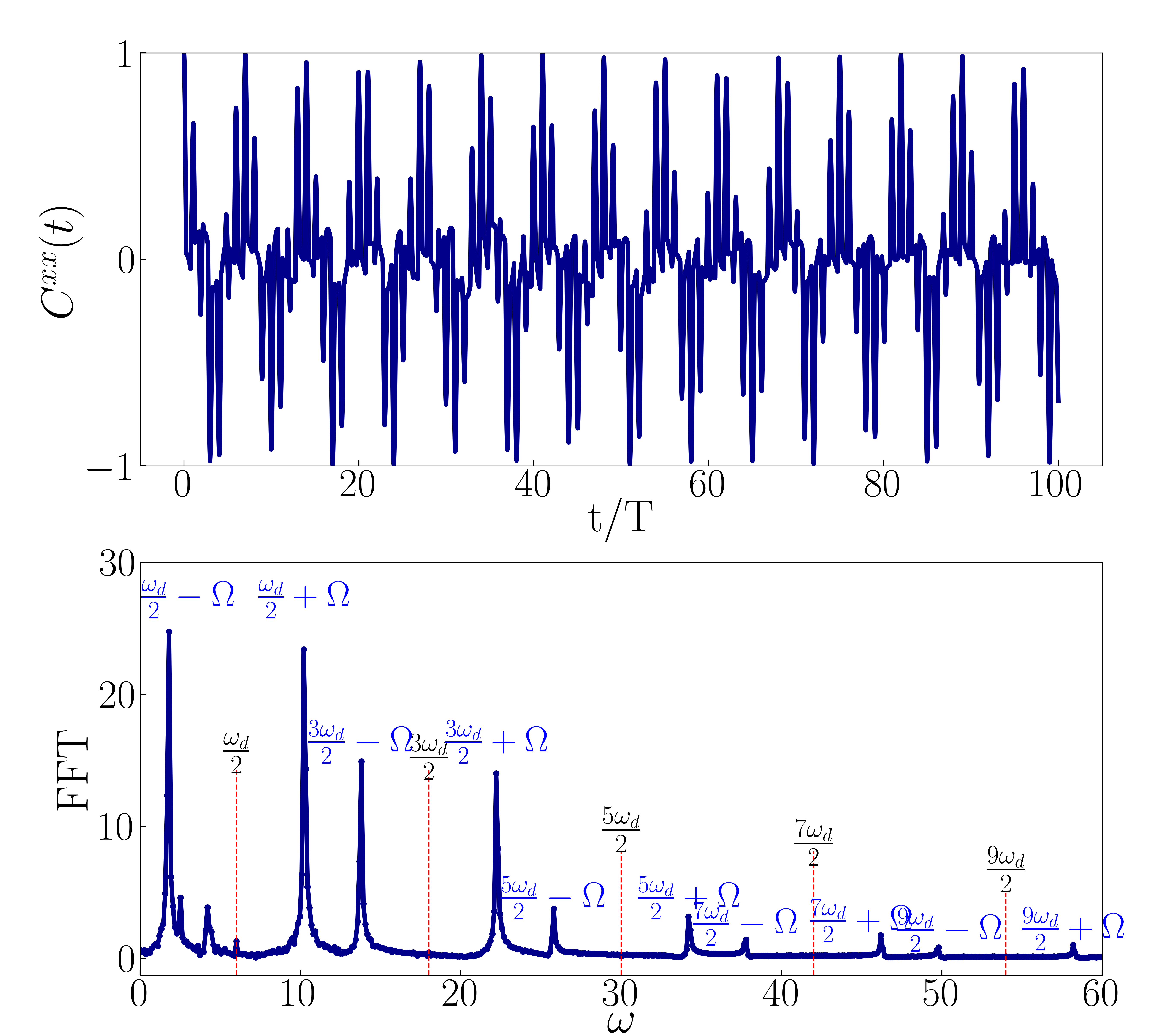}
	\centering
	\caption{\textbf{Top panel:} Time autocorrelation function $C^{xx}(t)$ for $J\ll h$ under a quasiperiodic drive. The correlations exhibit long-lived quasiperiodic oscillations for $\omega_d = 12$ and $L=6$, indicating robust non-equilibrium dynamical order.
		\textbf{Bottom panel:} Fast Fourier transform (FFT) of the time autocorrelation function(Dynamical structure factor). The corresponding Fourier spectrum displays peaks at frequencies $ k\omega_d/2 \pm \Omega $, with odd integer $k$, characteristic of subharmonic and quasiperiodic frequency mixing. For clarity of presentation, the spectral amplitudes are rescaled by a factor of $0.01$.}
	     \label{fig:C_tqc}
	\end{figure}

\begin{figure}
		\includegraphics[scale=0.23]{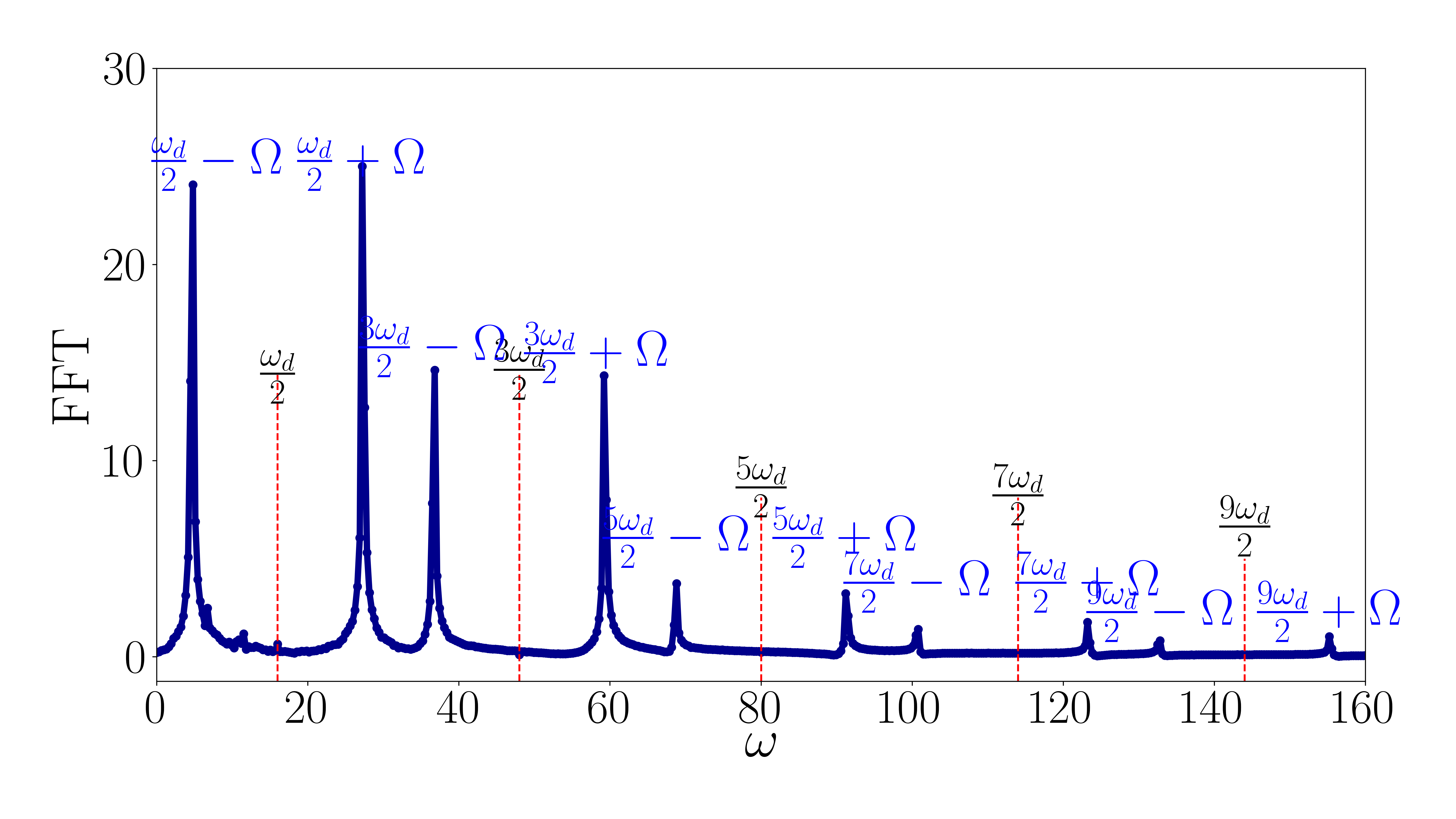}
		\centering
		\caption{FFT of the time autocorrelation function ($C^{xx}(t)$) for a chain of length $L = 6$ under quasiperiodic driving at $\omega_d = 32$. In the high-frequency regime, Floquet heating is exponentially suppressed in $\omega$, giving rise to a long-lived prethermal dynamical phase. Consequently, heating-induced sidebands are strongly reduced and the spectral weight becomes sharply concentrated around the subharmonic response peaks of the form $k\omega_d/2 \pm \Omega$, with odd integer $k$. The robustness of these locked frequency components signals exponentially long prethermal time scales and the stabilization of non-equilibrium order. The data are averaged over $200$ random realizations.}
		\label{fig:FFT_tqc}
	\end{figure}

	By contrast, when both driving frequencies are increased while keeping the irrational ratio $\Omega/\omega_d = \sqrt{2}/4$ fixed, the behavior of the autocorrelation function changes qualitatively. As demonstrated in Fig.~\ref{fig:C_tqc}, the oscillations of $C^{xx}(t)$ (similarly for $C^{yy}(t)$) become robust and persist over long times. The corresponding Fourier spectrum reveals peaks at frequencies $k\omega_d/2 \pm \Omega$, with  odd integer $k$. The appearance of these stable, incommensurate oscillations constitutes a clear signature of QTTSB, signaling the emergence of a TQC order. As observed in Fig.~\ref{fig:C_tqc}, several satellite peaks emerge around the primary subharmonic peak at $\omega_d/2-\Omega$. With increasing $\omega_d$ (while ratio $\Omega/\omega_d$ is fixed), these secondary peaks are gradually suppressed. As a result, the Fourier spectrum is dominated by a single primary subharmonic peak (see Fig.~\ref{fig:FFT_tqc}). This spectral purification clearly indicates the enhanced robustness and dynamical stability of the TQC phase in the high-frequency regime.
	
	The Fourier transform of the spin autocorrelation function defines the dynamical structure factor in the time domain,
	\begin{equation}
		S^{\alpha\alpha}(\omega) = \int dt \, e^{i\omega t} 
		\langle C^{\alpha\alpha}(t) \rangle .
	\end{equation}
		where $\alpha\in\{x,y,z\}$, in a TQC phase, $S(\omega)$ exhibits sharp Bragg-like peaks at discrete frequencies of the form
	\begin{equation}
		\omega = \sum_i n_i \omega_i ,
	\end{equation}
	where $\omega_i$ denote a set of fundamental frequencies that are incommensurate with each other and $n_i \in \mathbb{Z}$ \cite{Van Hove}.
	
\begin{figure}
   	\includegraphics[scale=0.18]{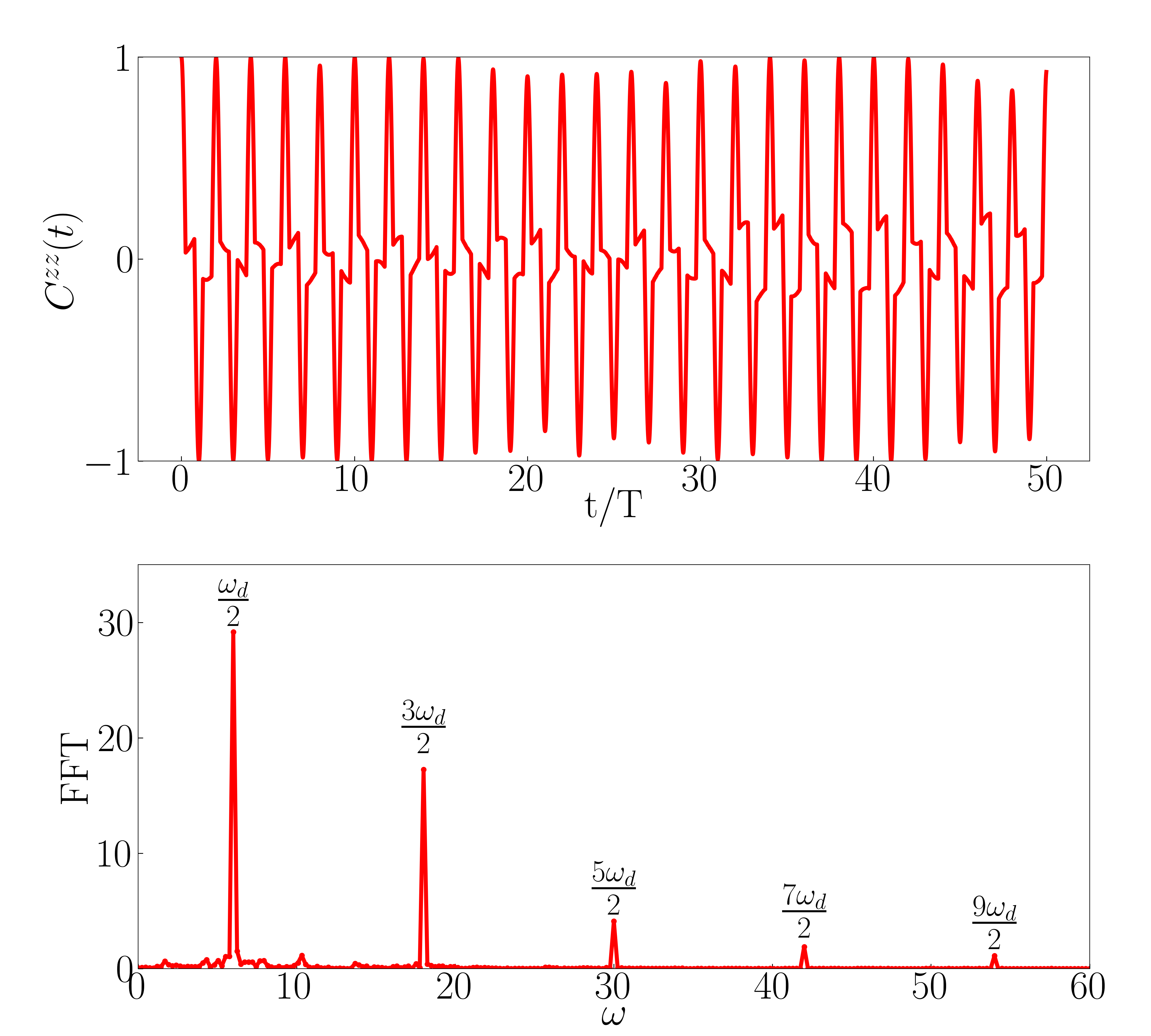}
   	\centering
   	\caption{\textbf{Top panel:} Time autocorrelation function $C^{zz}(t)$ in the $J\ll h$ regime under a quasiperiodic drive. The correlations exhibit long-lived quasiperiodic oscillations for $\omega_d = 12$ and $L=6$, indicating a robust non-equilibrium dynamical order. 
   		\textbf{Bottom panel:} FFT of the time autocorrelation function, $S^{zz}(\omega)$. The corresponding Fourier spectrum displays robust subharmonic peaks at frequencies $k\omega_d/2$ (where $k$ is an odd integer), confirming the breaking of quasi time-translation symmetry in the $z$-component. For clarity of presentation, the spectral amplitudes are rescaled by a factor of $0.01$.
   	}
   	\label{fig:Cz_tqc}
   \end{figure}

  While the initial state is polarized along the $x$-axis, when the time autocorrelation function is evaluated for the $z$-component, $C^{zz}(t)$ (see Fig.~\ref{fig:Cz_tqc}), the resulting Fourier spectrum $S^{zz}(\omega)$ displays a distinct behavior. Because the quasiperiodic drive primarily induces rotations in the $x-y$ plane, the frequency $\Omega$ of the longitudinal field does not explicitly appear in the position of the $z$-component's Fourier peaks. However, the spectrum still exhibits sharp subharmonic peaks at $k\omega_d/2$, where $k$ is an odd integer. Although the explicit dependence on $\Omega$ is absent due to the geometry of the drive, the presence of these robust subharmonic peaks at $k\omega_d/2$ demonstrates that the quasi time-translation symmetry is indeed broken in this measurement channel as well. Consequently, rather than just reflecting a direct response to the external drive, the autocorrelation $C^{zz}(t)$ effectively couples to the collective subharmonic dynamics, confirming that the TQC order is also manifested in the $z$-component.
	
	In a standard time crystal (TC), physical observables oscillate at frequencies distinct from the driving frequency, typically exhibiting a subharmonic response at $\omega = \omega_d/2$. In our model, the drive spectrum consists of components at $k\omega_d \pm \Omega$, and the TQC phase is characterized by observables responding at distinct subharmonic frequencies, specifically $k\omega_d/2 \pm \Omega$. Notably, although the dominant peaks of $S^{zz}(\omega)$ appear at $k\omega_d/2$ (multiples of half the primary drive frequency $\omega_d$), the system remains in a TQC phase. The absence of explicit $\pm\Omega$ shifts in the $z$-channel is a projection effect arising from the drive geometry for this specific observable. Nevertheless, the robust period-doubling observed in $S^{zz}(\omega)$ reflects the genuine subharmonic symmetry breaking characteristic of the underlying quasiperiodic state.

Crucially, the locations of these peaks are emergent and exhibit rigidity: under weak perturbations, the peak positions remain locked even though microscopic excitation energies may shift. This frequency locking is a manifestation of macroscopic temporal symmetry breaking rather than a consequence of fine-tuned microscopic parameters of the Hamiltonian.
	
	As illustrated in Fig.~\ref{fig:C_diff_omega}, the lifetime of this dynamical phase is highly sensitive to the driving frequency. For a small drive frequency ($\omega_d = 1$), the drive period is comparable to the local energy scales ($\omega_d \ll J$). Consequently, the system rapidly absorbs energy from the drive, leading to fast thermalization toward a featureless infinite-temperature state where quasiperiodic oscillations quickly decay. In stark contrast, for a larger frequency ($\omega_d = 12$), heating is strongly suppressed, and stable quasiperiodic oscillations emerge in the time autocorrelation function.

It is important to emphasize that the stability observed in the high-frequency regime has a fundamentally different physical origin from that arising in Floquet MBL. In Floquet MBL-protected time crystalline phases, the absence of thermalization is ensured by strong disorder and local integrals of motion. Here, stability instead stems from prethermalization due to the suppression of energy absorption at high driving frequencies.
	
	We have also analyzed the behavior of the transverse magnetization,
	\begin{equation}
		m^x(t) = \frac{1}{L} \sum_{i=1}^{L} \overline{\langle \hat{\sigma}_i^x(t) \rangle},
	\end{equation}
	which directly captures the collective spin dynamics along the transverse field direction. The overbar denotes the ensemble average over different random
	realizations and The $\langle \hat{\sigma}_i^x(t) \rangle$ indicates the expectation value computed over the fully time-evolved quantum state. 
	
\begin{figure}
		\includegraphics[scale=0.18]{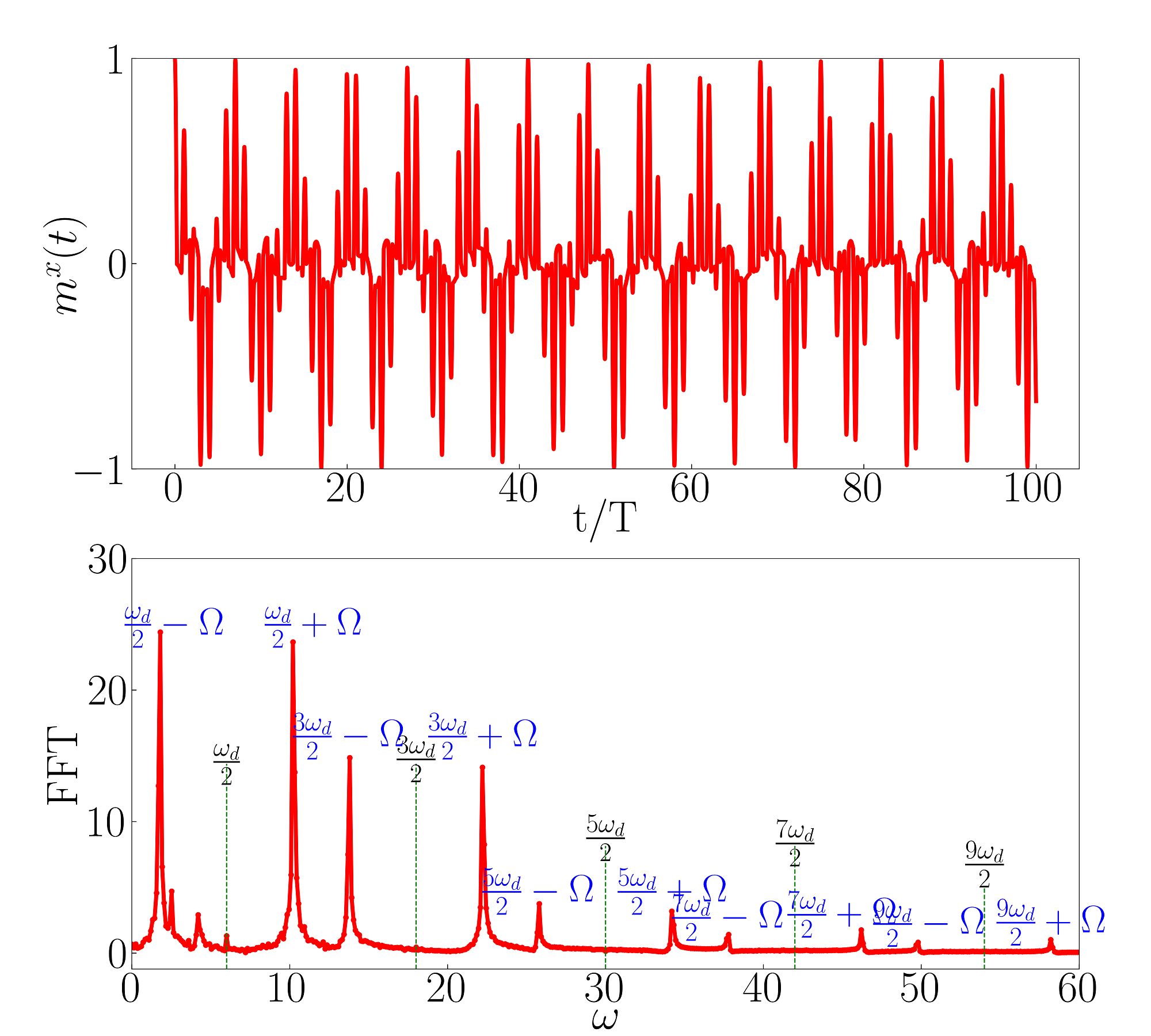}
		\centering
		\caption{\textbf{Top :} The transverse magnetization $m^x(t)$ in the $J\ll h$ regime under a quasiperiodic drive. The correlations exhibit long-lived quasiperiodic oscillations for $\omega_d = 12$ and $L=8$, indicating robust non-equilibrium dynamical order.
			\textbf{Bottom :} Fast Fourier transform (FFT) of the magnetization. The corresponding Fourier spectrum displays incommensurate peaks at frequencies $ k\omega_d/2 \pm \Omega $, with odd integer $k$, characteristic of subharmonic and quasiperiodic frequency mixing. For clarity of presentation, the spectral amplitudes are rescaled by a factor of $0.01$.}
		\label{fig:M_tqc}
	\end{figure}

	For $\omega_d = 12$ and $L=8$, the time evolution of $m^x(t)$ (as shown in Fig. \ref{fig:M_tqc} (top)) exhibits oscillatory behavior that maintains a clear quasiperiodic structure over the numerically accessible timescales. 
	
     We have also computed the FFT of the magnetization and display its amplitude in Fig.~\ref{fig:M_tqc} (Bottom). The resulting spectrum shows pronounced incommensurate peaks located at frequencies $k\omega_d/2 \pm \Omega$ (with odd $k$), consistent with the structure observed in the spin autocorrelation functions and the dynamical structure factor. The FFT of the magnetization, $\tilde{m}^x(\omega)$, directly probes the frequency content of the collective many-body response. The emergence of peaks at frequencies $k\omega_d/2 \pm \Omega$ reflects quasiperiodic frequency locking at the level of physical observables, independently confirming the structure identified in the autocorrelation function and dynamical structure factor.

\begin{figure}
		\includegraphics[scale=0.27]{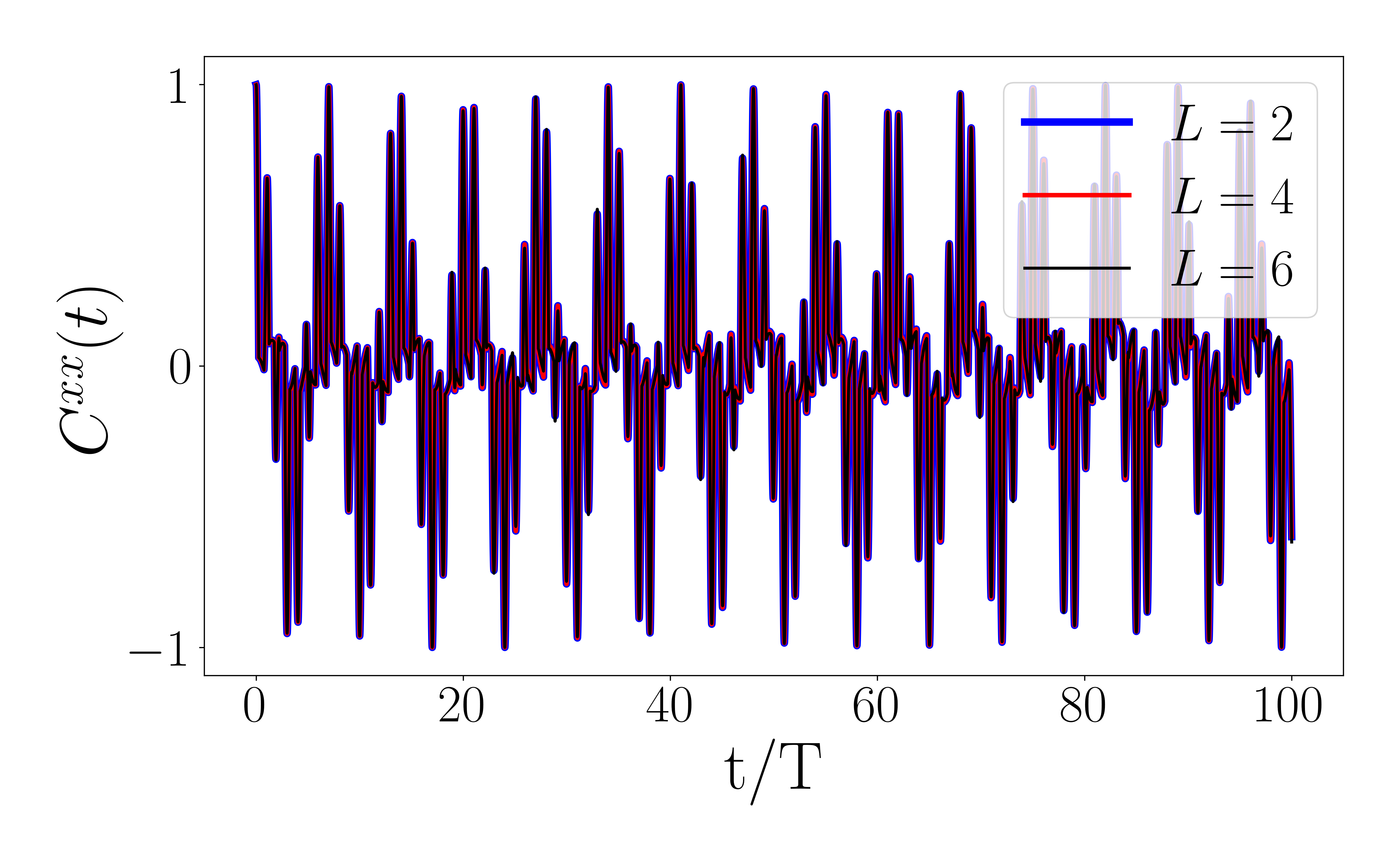}
		\centering
		\caption{ 
			Finite-size analysis of the temporal autocorrelation function ($C^{xx}(t)$) for system sizes $L=2,4,$ and $6$ under identical driving and disorder parameters. 
			Time evolution of the autocorrelation function shown well-defined quasiperiodic oscillations for all system sizes. 
			No noticeable change in the amplitude or qualitative structure of the oscillations is observed with increasing system size. 
			The temporal stability of these oscillations is primarily determined by the driving frequency $\omega$, indicating that the quasiperiodic response is governed by frequency hierarchy rather than global system size.}
		\label{fig:diff_size}
	\end{figure}

	We have further verified that the same oscillatory behavior and spectral peak structure persist for system size $L=8$, and the agreement with results for $L=6$ indicates that these spectral features are not finite-size artifacts but reflect intrinsic properties of the quasiperiodically driven spin chain system. To address possible finite-size effects, we analyzed the time autocorrelation function for system sizes $L = 2, 4,$ and $6$ using exact diagonalization under identical driving and disorder parameters. Figure~\ref{fig:diff_size} shows the time evolution of the autocorrelation function for these system sizes. In all cases, the dynamics exhibit well-defined quasiperiodic oscillations with identical frequency structure, and no noticeable qualitative change appears as the system size increases within the accessible range.
	
We observe that both the frequency content and amplitude structure of the quasiperiodic oscillations remain essentially unchanged with increasing system size. Within the studied range, we find no evidence that the observed TQC order is a finite-size artifact; rather, its stability is governed by frequency-controlled dynamical constraints, consistent with the general understanding of high-frequency driven many-body systems.

All reported results are averages over $200$ independent random realizations to suppress sample-to-sample fluctuations. The time evolution is computed with a small time step of $T/200$, where $T$ is the driving period, providing adequate resolution to capture continuous dynamics and suppress numerical discretization errors.

In summary, these results indicate that the observed time-periodic order arises not from actual localization but via a prethermal mechanism stabilized by the drive frequency. This behavior can therefore be classified as prethermal TQC, where quasiperiodic time-translation symmetry breaking (QTTSB) remains stable for long yet finite times.

In the prethermal regime, thermalization is not fully prohibited but is delayed for much longer times, as energy absorption from the driving field is exponentially suppressed at high frequencies ($\omega_d \gg J$). Consequently, the system enters a metastable, long-lived phase during which discrete or quasiperiodic time-translation symmetry is effectively broken, though eventual thermal equilibrium is approached after a very long time. This long-lived regime is referred to as the prethermal time, which we define as the duration of the first plateau in the system's entanglement entropy (further discussed in the entanglement entropy section). Notably, the prethermalization time depends strongly on the driving protocol. For periodically driven systems, it grows exponentially with driving frequency: $\tau_{\text{pre}} \propto \exp(c\omega_d)$ \cite{Pragna Das}. For systems driven by the Thue–Morse sequence, sub-exponential growth ($\tau_{\text{pre}} \propto \exp(C [\ln(\omega_d/g)]^2)$ with constant $g$) \cite{Takashi Mori} and stretched-exponential growth ($\tau_{\text{pre}} \propto \exp(C \omega_d^\beta)$ with $0<\beta<1$) \cite{Pragna Das} have been reported. Meanwhile, for random multipolar driving (RMD), the prethermal time exhibits power-law scaling: $\tau_{\text{pre}} \propto \omega_d^{2n+1}$ for $n\ge1$ \cite{r35}.

	\section{Entanglement entropy}\label{Sec:entanglement entropy}
	Entanglement entropy (EE) $S$ is a standard measure of bipartite entanglement. It is defined as the von Neumann entropy of the reduced density matrix of a subsystem A ($\rho_A=Tr_B\rho$), under the condition that the global system is in a pure state $\psi$ ($\rho=|\psi\rangle\langle\psi|$). The EE,
	$
	S_A = -\mathrm{Tr}_A (\rho_A \ln \rho_A),
	$
vanishes for unentangled product states and reaches a maximum for maximally entangled states \cite{Calabrese2004,Latorre2004,Igloi2007}.  
	
To characterize the growth of quantum correlations, we study the time evolution of the half-chain EE, defined as
		\begin{equation}
		S(t) = - \mathrm{Tr} \, \rho_{L/2}(t) \ln \rho_{L/2}(t),
	\end{equation}
	where $\rho_{L/2}(t)$ is the reduced density matrix of half of the chain,
	\begin{equation}
		\rho_{L/2}(t) = \mathrm{Tr}_{1 \le i \le L/2} \, |\psi(t)\rangle \langle \psi(t)|,
		\label{eq.entropy}
	\end{equation}
	and $|\psi(t)\rangle$ is the evolved state of the system at time $t$. 
	In generic non-integrable many-body systems, highly excited eigenstates obey the ETH, and the EE exhibits volume-law scaling. Eigenstates near the spectrum center correspond effectively to infinite temperature, with entanglement properties well approximated by random pure states. Their EE thus approaches the Page value $S_{\text{Page}}$, close to the maximum possible entropy $S_{\text{max}}$. For a one-dimensional spin-$1/2$ chain of length $L$ bipartitioned at $L_A = L/2$,
\begin{equation}
    S_{\text{Page}} = \frac{L \ln 2 - 1}{2}, \quad S_{\text{max}} = \frac{L \ln 2}{2},
\end{equation}
indicating eventual thermalization \cite{Page1993,Zhao2021}.

Here, we examine how quasiperiodic modulation affects the half-chain EE $S(t)$ across different driving frequencies.
		\begin{figure}
		\includegraphics[scale=0.28]{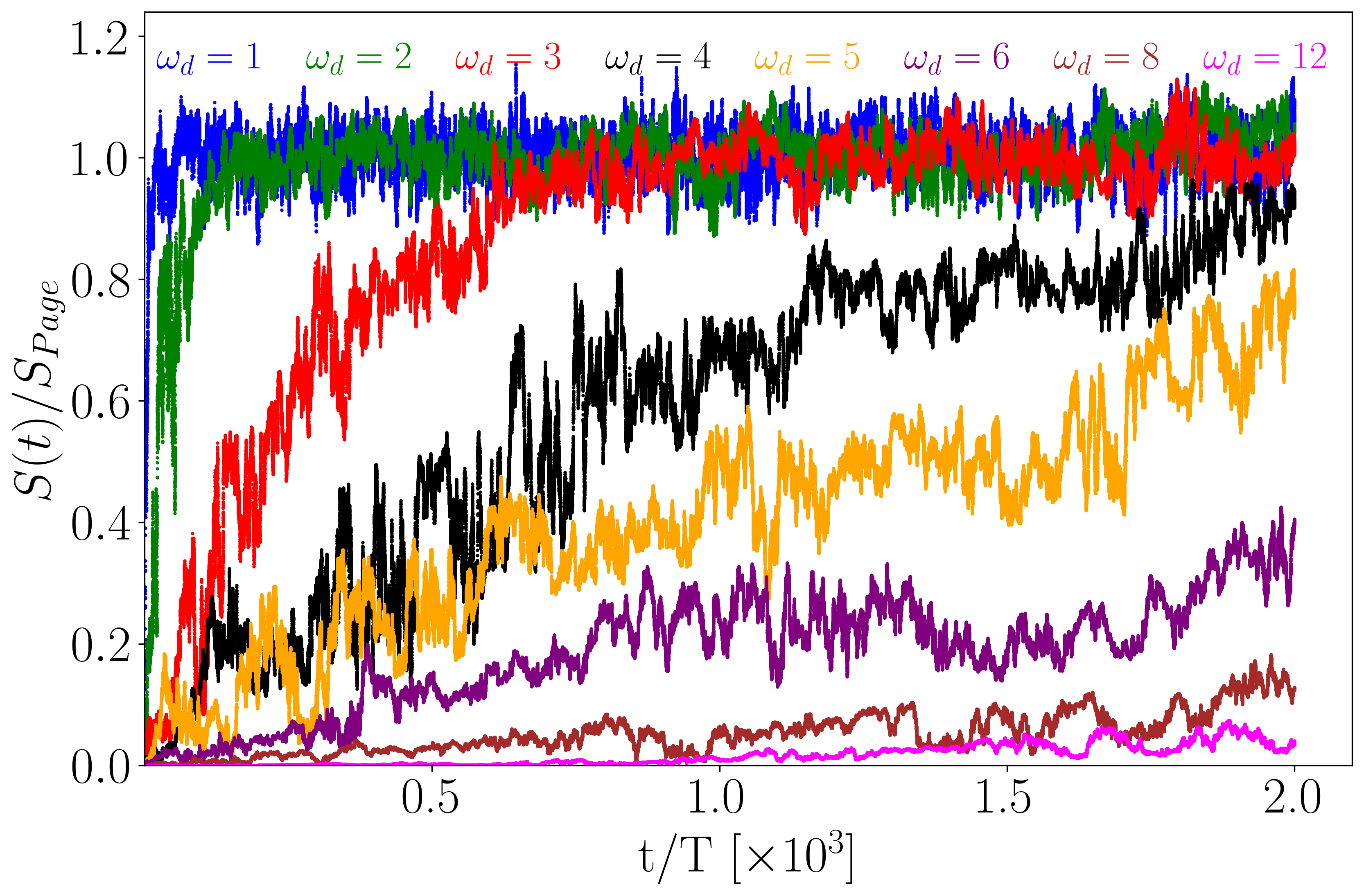}
		\centering
		\caption{Temporal growth of the entanglement entropy for different values of drive frequencies. $\Omega/\omega=\frac{\sqrt{2}}{4}$, $J=5.5$, $h=0.3$ and $L=6$, increasing the absolute driving frequencies slows the buildup of EE, reflecting the suppression of energy absorption and heating in the high-frequency regime.}
		\label{fig:bipartite}
	\end{figure}
	
	For a chain of length $L = 6$ with periodic boundary conditions, we employ exact diagonalization to compute the real-time dynamics generated by the Hamiltonian in Eq.~\eqref{eq:H_total}, the time evolution is initialized from the product state
	\begin{equation}
		|\psi_0\rangle = \bigotimes_{i=1}^L 
		\frac{|\uparrow_i\rangle + |\downarrow_i\rangle}{\sqrt{2}},
	\end{equation}
	which corresponds to a fully polarized state along the transverse fields direction and contains no initial entanglement.
	
	Averaging over $200$ disorder realizations, we compute $S(t)$ up to times $2\times 10^3T$ for different values of $\omega_d$, while keeping the ratio $\Omega/\omega_d=\sqrt{2}/4$. The results are presented in Fig.~\ref{fig:bipartite}.
	
	At low driving frequencies, the entanglement entropy exhibits a rapid temporal growth followed by a quick saturation to the Page value expected for thermal states. Such fast entanglement production signals strong energy absorption from the drive and thermalization. Consequently, any emergent temporal order, including a prethermal TQC order, is destroyed on short time scales, indicating that the prethermal window is either absent or extremely narrow. This behavior is consistent with general expectations for interacting driven systems, where entanglement typically grows linearly at early times before saturating due to heating and thermalization \cite{DAlessio2014,Igloi2007,Abanin2015}.
	
	In contrast, upon increasing the absolute values of the driving frequencies while keeping $\Omega/\omega_d = \sqrt{2}/4$ fixed, the temporal growth of entanglement entropy becomes significantly slower. This slowdown reflects the suppression of resonant energy absorption processes in the high-frequency regime. In this regime, the dynamics remain remarkably stable over parametrically long times, resulting in an extended non‑equilibrium plateau in $S(t)$. The reduced rate of entanglement spreading thus serves as a clear dynamical signature of suppressed heating and the stabilization of a long‑lived prethermal regime before eventual thermalization. 
	\begin{figure}[htb]
	\centering
	\includegraphics[width=0.45\textwidth]{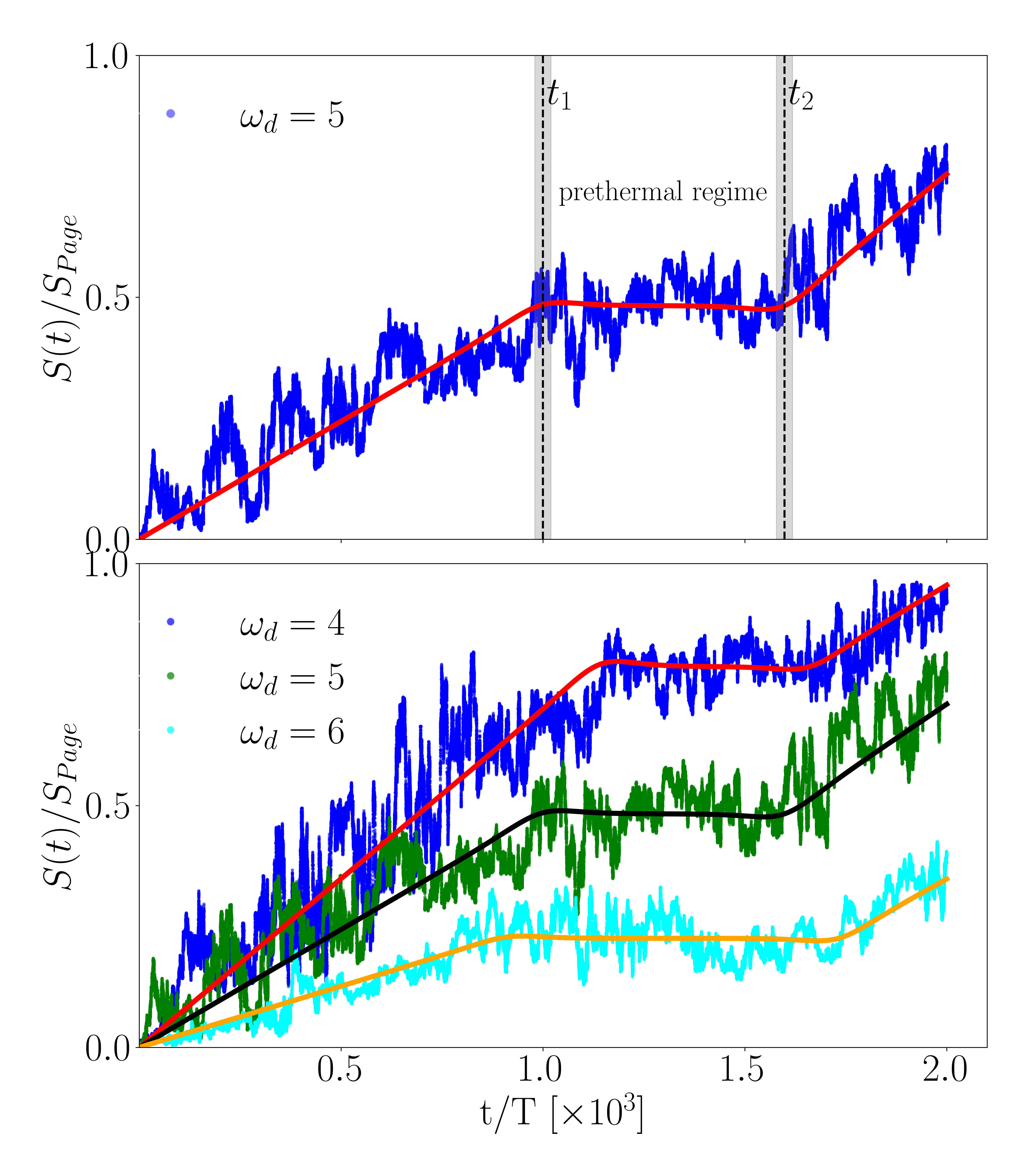}
	\caption{Temporal growth of the EE $S(t)/S_{\text{Page}}$, illustrating the emergence and frequency dependence of the prethermal regime. \textbf{(Top)} For $\omega_d = 5$, a distinct prethermal plateau appears between $t_1$ and $t_2$, characterized by suppressed entropy growth. The red line is a guide to the fit; dashed shaded areas indicate uncertainty in the plateau boundaries. \textbf{(Bottom)} EE for driving frequencies $\omega_d = 4, 5, 6$. As $\omega_d$ increases, the plateau duration lengthens significantly, reflecting the suppression of resonant energy absorption at high frequencies, delayed thermalization, and the stabilization of prethermal dynamics over parametrically long timescales. Parameters: $\Omega/\omega_d = \sqrt{2}/4$, $J = 5.5$, $h = 0.3$, $L = 6$.}
	\label{fig:prethermal_combined}
\end{figure}

	Importantly, the emergence of a long-lived entanglement plateau (highlighted in Fig.~\ref{fig:prethermal_combined} (top)) is directly tied to the stability of the prethermal TQC. Since the underlying temporal order depends on the persistence of nonthermal dynamics, the suppression of entropy growth indicates that the system remains governed by an effective quasi-conserved description over extended timescales, meaning that its dynamical features resemble those of a conserved system, even though such characteristics are only apparent within specific time frames and under certain conditions. As the driving frequency increases, the lifetime of this prethermal regime ($\tau_{pre}=t_2-t_1$) is parametrically enhanced, thereby strengthening the robustness of the TQC order against heating-induced decay. In contrast, when the entanglement entropy rapidly saturate to the Page value, the duration of the plateau tends to zero, and the associated temporal order quickly destabilizes. Entanglement dynamics therefore provide not only a probe of heating, but also a quantitative measure of the stability and lifetime of the prethermal TQC phase.

	We further emphasize that the lifetime of the prethermal regime exhibits a strong dependence on the driving frequency. As illustrated in Fig.~\ref{fig:prethermal_combined} (bottom), increasing the value of the drive frequency leads to a pronounced extension of the prethermal plateau. In other words, the duration over which the system remains trapped in the nonthermal regime grows significantly with increasing frequency. This behavior is consistent with the general expectation that high-frequency drive suppresses resonant energy absorption processes, thereby delaying thermalization and stabilizing the effective prethermal description over parametrically long time scales. Specifically, at high frequencies, the system lacks sufficient time to respond to the drive fields, resulting in minimal energy absorption. Conversely, at lower frequencies, the system has adequate time to absorb energy, ultimately facilitating thermalization.
	
\begin{figure}
		\includegraphics[scale=0.28]{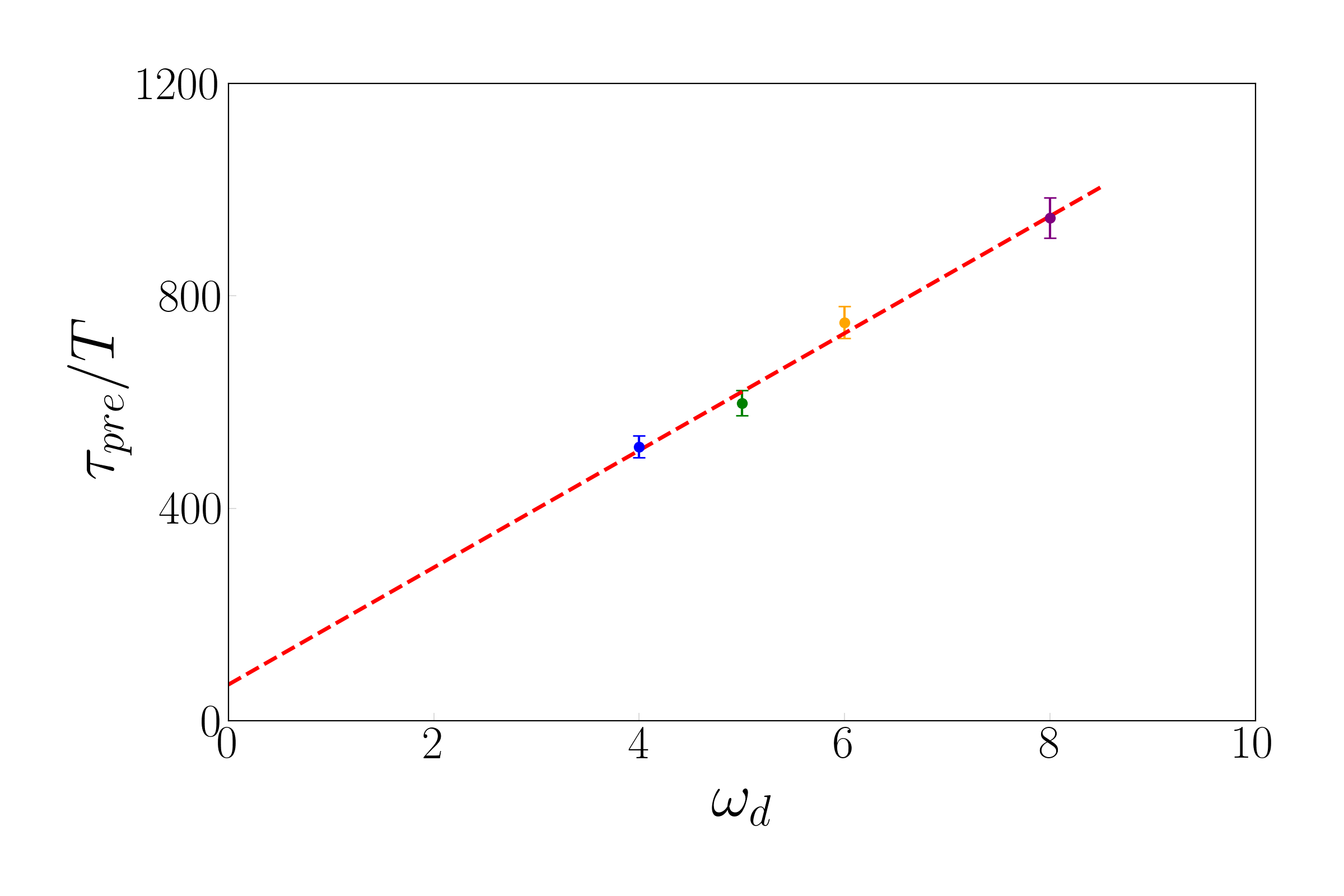}
		\centering
		\caption{The prethermal time as a function of omega shows that with increasing omega, the prethermal time significantly increases. In this graph, the points representing the prethermal time have an error . Although this graph does not provide the precise and parametric behavior of the prethermal time in terms of omega, it clearly demonstrates the increase in prethermal time length with increasing omega.}
		\label{fig:time_Vs_omega}
	\end{figure}

	In Fig.~\eqref{fig:time_Vs_omega}, we plotted the $\tau_{pre}$ as a function of $\omega_d$, which shows that the prethermal time increases with increasing omega. Although our data cannot precisely capture the exact behavior of the prethermal time as a function of $\omega$, it clearly indicates that the prethermal time grows as $\omega$ increases.
	
	\subsection{Impact of Disorder Sampling Protocols on Thermalization Dynamics}
	
	To investigate the role of disorder and its chosen sampling interval on the system's dynamics, we compared the behavior of EE under two different distributions. In this analysis, the driving frequency is set to a low value, $\omega_d=1$, and $\Omega/\omega=\sqrt{2}/4$. The first interval represents a symmetric distribution in the range $[-J/2, J/2]$, while the second is an asymmetric distribution in the range $[J/2, 3J/2]$.
	
\begin{figure}[htbp]
		\centering
		\includegraphics[scale=0.28]{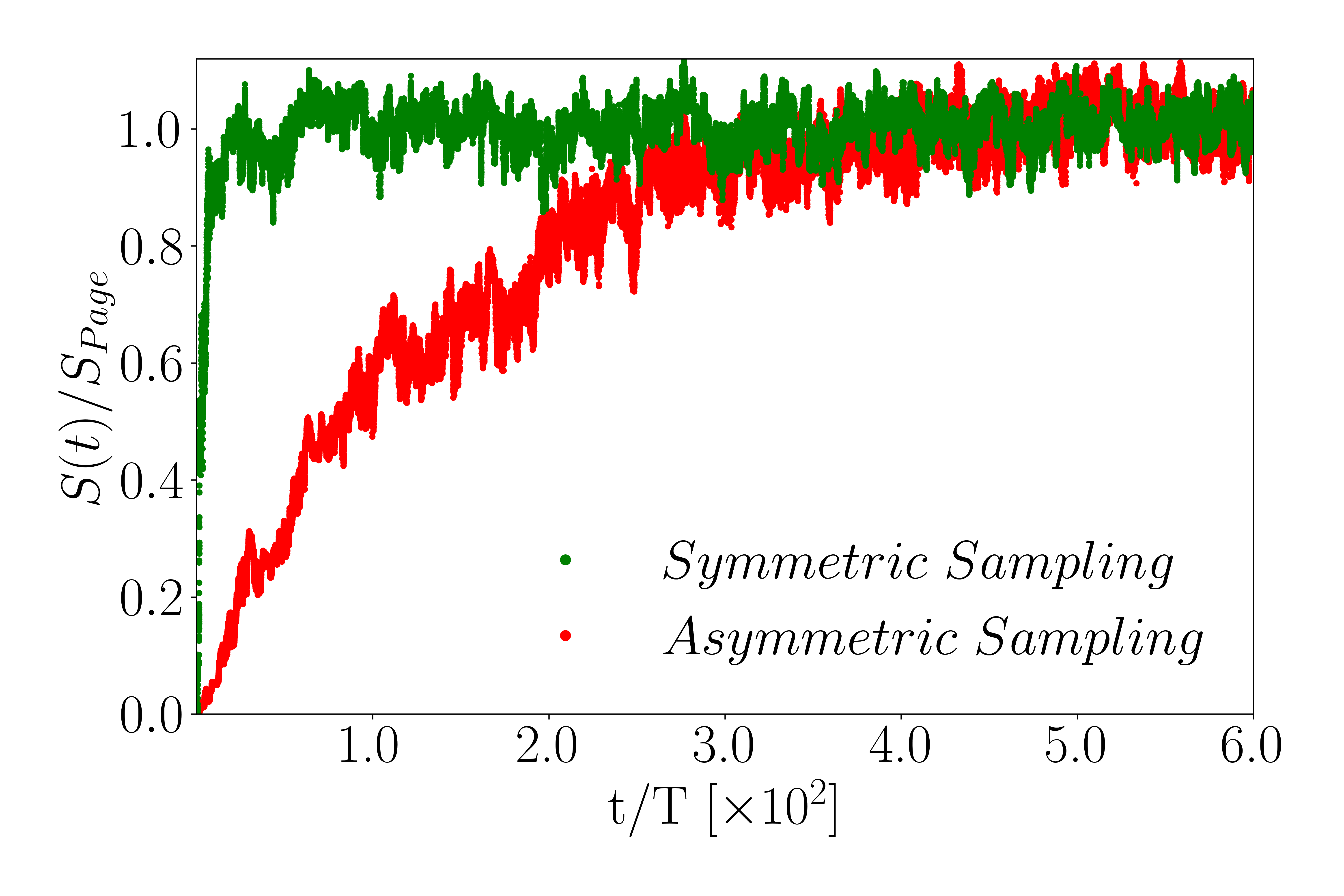}
		\caption{Temporal growth of the EE at a low driving frequency ($\omega_d=1$ and $\Omega/\omega_d = \sqrt{2}/4$) for two distinct disorder sampling protocols. The green curve corresponds to a symmetric disorder distribution sampled from $[-J/2, J/2]$, exhibiting rapid entropy growth indicative of fast thermalization. In contrast, the red curve illustrates the dynamics for an asymmetric distribution sampled from $[J/2, 3J/2]$, where the non-zero mean establishes a robust macroscopic order that strongly suppresses energy absorption and slows down thermalization.}
		\label{fig:entropy_sampling}
	\end{figure}

	As shown in the Fig.~\ref{fig:entropy_sampling}, the choice of these intervals has a striking impact on the entropy growth rate. In the symmetric distribution case (green curve), the entropy grows rapidly, indicating quick thermalization of the system. Conversely, in the asymmetric distribution (red curve), the entropy growth is considerably slower, and the system exhibits resistance to heating.
	
The physical origin of this behavior stems from the distinct nature of the system's phase in these two regimes. For the asymmetric interval, the system resides in a globally ordered (ferromagnetic) phase. In this regime, strong effective spin-spin coupling provides collective rigidity, rendering the system highly resistant to drive-induced excitations. Conversely, for the symmetric interval centered around zero, the system lies in a highly random, disordered phase. Lacking protective macroscopic order, the spins are readily excited even by weak drives, losing their local structure and leading to rapid energy absorption and thermalization.

\section{Stability of the TQC Phase }\label{Sec:Rigidity}
  
  In this section, we examine the stability of the TQC phase under various perturbations. As a preliminary step, it is instructive to review the general mechanisms by which nonequilibrium quantum many-body systems absorb energy and thermalize under different classes of driving fields.
  
  In systems governed by time-dependent local Hamiltonians, energy absorption and thermalization occur through local transitions, such as localized spin flips or the creation of short-range excitations, induced by an external drive. In strongly disordered systems, these local processes require a specific, finite energy cost. When subjected to a periodic drive featuring a discrete frequency spectrum, these local energy gaps are typically off-resonant with the periodic drive. This mismatch suppresses energy absorption, allowing the MBL phase to remain stable. In contrast, random driving protocols exhibit a continuous frequency spectrum that invariably contains the exact frequencies required to resonantly activate arbitrary local transitions. Consequently, random driving generically leads to unbounded energy absorption, thereby destabilizing the MBL phase and driving the system toward thermalization \cite{P. T. Dumitrescu,Zhao2021}.
  
  Quasiperiodic driving occupies an intermediate regime between periodic and random drives. Although it consists of a discrete set of incommensurate frequencies, the resulting frequency spectrum is dense rather than continuous. Under typical conditions, this structure can strongly suppress resonant processes and thus stabilize over long times. Nevertheless, at asymptotically long times, rare resonances can accumulate and eventually drive the system toward thermalization. Importantly, the associated heating timescale diverges exponentially for weak driving amplitudes or high drive frequencies, leading to an extended regime of slow relaxation \cite{P. T. Dumitrescu,Zhao2021}.
  
  The existence of such a long-lived quasi-localized regime suggests the emergence of transient dynamical phases that are unique to quasiperiodically driven systems. In close analogy with prethermal states in periodically driven systems, this regime can be naturally interpreted as a prethermal TQC phase. While the system ultimately thermalizes to an infinite-temperature state, the thermalization process is sufficiently slow to permit the robust observation of quasiperiodic temporal order over experimentally and numerically relevant timescales.
  The stability of this prethermal regime exhibits a pronounced sensitivity to variations in several control parameters, with the drive frequency being particularly noteworthy. Specifically, increasing the drive frequency serves to substantially enhance the longevity of the TQC phase. Moreover, the heating dynamics are strongly influenced by the smoothness of the driving protocol: non-smooth drives, such as rectangular pulses, enhance high-frequency spectral components and significantly accelerate energy absorption, whereas smooth drives (e.g., sinusoidal modulation) markedly suppress heating and extend the prethermal window. The drive amplitude also plays a crucial role, with stronger driving leading to faster thermalization \cite{P. T. Dumitrescu,Zhao2021}.
  
  In the following, we systematically analyze the stability of the TQC phase, with particular emphasis on the role of the NNN couplings as a perturbation and imperfections rotation, and identify the parameter regimes in which prethermal quasiperiodic order remains robust.
	
\subsection{Stability of the TQC Phase Under The  NNN couplings Perturbations}

To assess the stability of the TQC phase against weak interactions, we introduce a perturbation term to the Ising transverse field (ITF) Hamiltonian from Eq. (\ref{eq:H_ITF}):

\begin{equation}
	\hat{H}_{\text{ITF}}(t)=h_1(t)\sum\limits_{i}\left[J_{i}\hat{\sigma}_{i}^{z}\hat{\sigma}_{i+1}^{z} + h_{i}\hat{\sigma}_{i}^{x} + J_2 \hat{\sigma}_{i}^{z}\hat{\sigma}_{i+2}^{z}\right].
	\label{Eq:ITF-Hamiltonian-quasi-int}
\end{equation}

Here, $J_2$ controls the strength of the NNN $zz$-coupling. This coupling acts as a perturbation that can lead to system heating. The time-dependent coefficient $h_1(t)$, which is periodically switched on and off according to Eq. (\ref{Eq:h1-h2}), ensures this perturbation is applied only during specific intervals, mimicking a pulsed interaction.

The stability of the TQC phase is probed by analyzing the FFT (or dynamical structure factor) of the time autocorrelation function $C^{xx}(t)$. Prominent peaks at frequencies $k\omega_d/2 \pm\Omega$ (with odd integer $k$) persist for finite $J_2$, confirming the robustness of the TQC phase over a range of the nearest-neighbor coupling strengths.
\begin{figure}[htbp]
	\centering
	\includegraphics[scale=0.18]{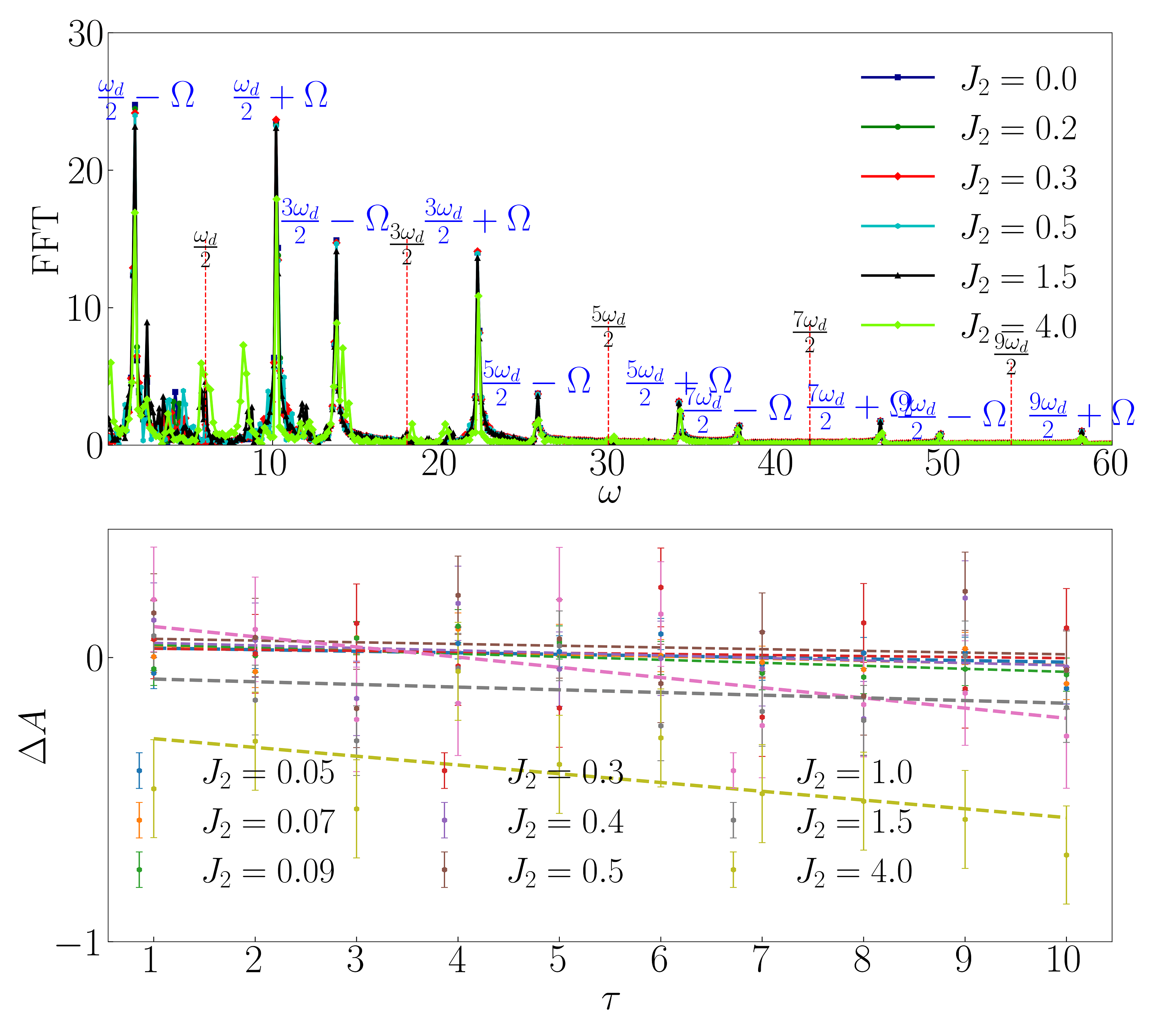}
	\caption{
		\textbf{Top:} FFT of $C^{xx}(t)$ in the TQC phase for varying the NNN coupling strength $J_2$ for $\omega_d=12$ and $\Omega/\omega_d=\sqrt{2}/4$. Data are obtained from exact diagonalization of a disordered ITF chain  of length $L=6$, in $J\ll h$ regime, averaged over $200$ disorder realizations. The FFT is computed over $100T$. Peaks at $k\omega_d/2 \pm\Omega$ ($k$ is  an odd integer) indicate TQC stability. Peak amplitudes decrease with increasing $J_2$, vanishing at a critical value ${J_2}^c\ll 1$.
		\textbf{Bottom:} Deviation $\Delta A(\tau) = A_{J_2}(\tau) - A_{J_2=0}(\tau)$ of the first-peak amplitude across different time windows $\tau$. For $L=6$, the TQC phase becomes unstable for $J_2 > {J_2}^c$, evidenced by the decay of $\Delta A$ for $J_2 > {J_2}^c$. Error bars represent the standard deviation across time windows.
	}
	\label{Fig:FFT-quasi-int}
\end{figure}

Two main features emerge from the FFT. For $J_2 < {J_2}^c$, the spectral weight remains sharply concentrated around the subharmonic frequencies of the form $k\omega_d/2 \pm \Omega$, with odd integer $k$, indicating that the quasiperiodic frequency locking of the TQC phase is robust in the weak-interaction regime. 

As $J_2$ increases toward $ {J_2}^c$, additional peaks beyond the primary subharmonic components gradually develop in the spectrum. These extra spectral features signal the onset of the NNN coupling-induced mode mixing and the progressive destabilization of the locked quasiperiodic response. 

For sufficiently strong the NNN couplings, redistribution of spectral weight away from the subharmonic structure reflects the breakdown of the TQC phase. At the critical strength ${J_2}^c$, the subharmonic response is no longer dominant, marking the loss of long-lived non-equilibrium order.
The absence of significant spectral weight at other frequencies implies that the perturbation does not excite new resonant modes but merely attenuates the existing TQC response. The amplitude eventually vanishes at the critical $ {J_2}^c$, signaling the breakdown of the TQC phase(see up panel of Fig. \ref{Fig:FFT-quasi-int}).
Second, the subharmonic peak amplitude remains nearly constant within a finite range of $J_2 < {J_2}^c$, indicating the stability of the locked quasiperiodic response. However, as $J_2$ exceeds the critical value ${J_2}^c$, the FFT amplitude progressively decreases, reflecting the weakening of the subharmonic order and the eventual destabilization of the TQC phase.

We quantify $ {J_2}^c$ using a windowed Fourier analysis. Dividing the total time span ($100T$) into ten segments, we compute the amplitude $A_{J_2}(\tau)$ of the first dominant peak within each time window $\tau$. The deviation $\Delta A(\tau) = A_{J_2}(\tau) - A_{J_2=0}(\tau)$ measures the nearest-neighbor coupling-induced suppression. As shown in the bottom panel of Fig. \ref{Fig:FFT-quasi-int}, $\Delta A$ remains near zero for $J_2 < {J_2}^c$ but decays for $J_2 > {J_2}^c$. For a system of length $L=6$, this analysis yields $ 0.5\ll{J_2}^c \ll 1.0$.

\subsection{Stability of the TQC Phase Under Imperfect Rotations}

We next investigate the effect of imperfect spin rotations, modeled by introducing a deviation parameter $\epsilon$ into the kick Hamiltonian of Eq. (\ref{eq:H_drive}):
\begin{equation}
	\hat{H}_{\text{Kick}}(t)=h_2(t)\sum\limits_{i}\left[(1-\epsilon)\cos(\Omega t)\hat{\sigma}_{i}^{y} - \sin(\Omega t)\hat{\sigma}_{i}^{x}\right].
	\label{Eq:quasi-Hamiltonian_drive_imp}
\end{equation}
The factor $(1-\epsilon)$ attenuates the $y$-component of the rotation, capturing a common control of imperfection.

\begin{figure}[htbp]
	\centering
	\includegraphics[scale=0.18]{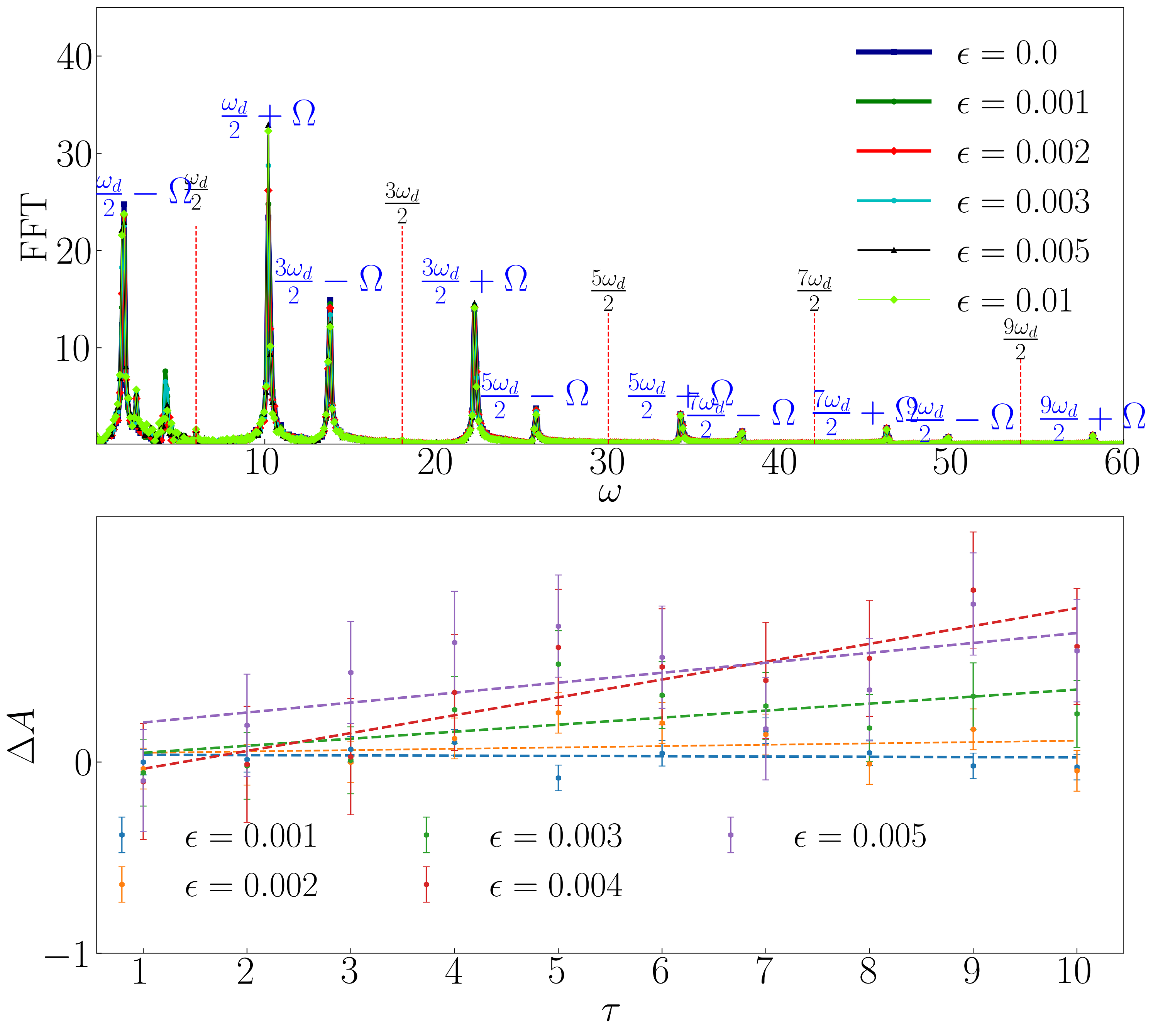}
	\caption{
		\textbf{Top:} FFT magnitude of $C^{xx}(t)$ for varying rotation imperfection $\epsilon$ for $\omega=12$ and $\Omega/\omega_d=\sqrt{2}/4$. Parameters and averaging are identical to Fig. \ref{Fig:FFT-quasi-int}. Peaks at $k\omega_d/2 \pm\Omega$ demonstrate TQC resilience. Amplitude suppression leads to vanishing peaks at a critical $\epsilon_c \sim 0.002$ for $L=6$.
		\textbf{Bottom:} Deviation $\Delta A(\tau) = A_{\epsilon}(\tau) - A_{\epsilon=0}(\tau)$ of the first-peak amplitude across time windows. For $\epsilon > \epsilon_c \sim 0.002$, $\Delta A$ increase.
	}
	\label{Fig:FFT-quasi-imperfect}
\end{figure}

The FFT of $C^{xx}(t)$ for various $\epsilon$ is shown in Fig. \ref{Fig:FFT-quasi-imperfect}. The impact of this imperfection is qualitatively different from the NNN coupling : the characteristic peaks at $k\omega_d/2 \pm\Omega$ ($k$ odd) retain their positions but their amplitudes are increased as $\epsilon$ increases.

For imperfection below a critical threshold $\epsilon_c$, the FFT remains sharply concentrated around the subharmonic frequencies of the form $k\omega_d/2 \pm \Omega$, indicating that the essential frequency-locked structure of the TQC is preserved. As the imperfection $\epsilon$ increases beyond $\epsilon_c$, additional spectral peaks don't emerge in the FFT and the peak positions are fixed. But the peak value are increased due to increasing of $\epsilon$.  In other words, this imperfection does not introduce new incommensurate driving frequencies, thereby preserving the fundamental frequency-locking mechanism of the TQC phase. Consequently, the peak positions remain fixed at $k\omega_d/2 \pm \Omega$. However, the imperfection geometrically breaks the circular symmetry of the ideal drive in the $xy$-plane, transforming it into an elliptical drive where the $y$-component is attenuated. Because our tracked observable is the $x$-component correlation function, $C^{xx}(t)$, this elliptical squeezing effectively increases the relative projection of the spin trajectories along the $x$-axis. As a result, the spectral weight of the response in the $x$-channel is enhanced, leading to an increase in the amplitude of the corresponding peaks without altering their frequencies.

We determine the critical imperfection $\epsilon_c$ using the same windowed amplitude analysis. The deviation $\Delta A(\tau) = A_{\epsilon}(\tau) - A_{\epsilon=0}(\tau)$ is plotted in the bottom panel of Fig. \ref{Fig:FFT-quasi-imperfect}. For a chain of length $L=6$, $\Delta A$ remains stable for $\epsilon < \epsilon_c$ and increases for $\epsilon > \epsilon_c$, yielding $\epsilon_c \approx 0.002$.

In summary, in contrast to the $J_2$ perturbation, increasing the rotation imperfection $\epsilon$ yields an initial, counterintuitive increase in the amplitude of the primary peaks at $k\omega_d/2 \pm \Omega$. It is crucial to note that this enhancement does not imply an increased stability of the TQC phase against heating. Rather, it arises from a geometric effect: the imperfection $\epsilon$ does not introduce new incommensurate driving frequencies, thereby preserving the frequency-locking mechanism. However, it geometrically breaks the circular symmetry of the ideal drive in the $xy$-plane, transforming it into an elliptical drive where the $y$-component is attenuated. Because our tracked observable is the $x$-component correlation function, $C^{xx}(t)$, this elliptical squeezing effectively increases the relative projection of the spin trajectories along the $x$-axis. As a result, the spectral weight in the $x$-channel is enhanced without altering the peak frequencies. Nevertheless, this geometric enhancement is bounded; for $\epsilon > \epsilon_c$ (where $\epsilon_c \sim 0.002$ for $L=6$), the imperfection becomes severe enough to disrupt the precise spin-flip resonance required for the prethermal regime, leading to rapid heating and the ultimate destruction of the TQC phase.

Furthermore, the TQC phase exhibits finite resilience against both the NNN coupling and imperfection, provided $J_2<J_2^c$, and $\epsilon<\epsilon_c$. This robustness is crucial for potential experimental realizations where such imperfections are inevitable.

It is intuitive to compare the stability characteristics of prethermal TQCs with  conventional TCs, We observe that the critical perturbation thresholds for maintaining stability in prethermal TQCs are roughly an order of magnitude lower than those for conventional TCs . This reduced robustness stems from the more intricate synchronization mechanisms present in quasiperiodically driven systems. While periodic drives can induce robust quantum frequency locking \cite{r27}, whereby the system locks onto rational multiples of the driving frequency, synchronization under incommensurate drives involves more subtle effects.  

In quasiperiodically driven systems, the response can lock onto linear combinations of the incommensurate driving frequencies, producing a frequency spectrum with peaks at $m\omega + n\Omega$ with integers $m$ and $n$. This irrational mode locking \cite{marripour}, analogous to phenomena in quasiperiodic spatial structures, underlies the quasi-stable, aperiodic oscillations observed in TQCs. Although this mechanism enriches the spectral response, it also renders the phase more susceptible to perturbations than the simpler, robust locking seen in conventional TCs.  

Furthermore, our data indicate that the choice of interaction strengths $J_i$ significantly affects stability. When $J_i$ is chosen symmetrically, the prethermal TQC exhibits weaker stability compared to an antisymmetric choice (e.g., $\{J/2, 3J/2\}$) corresponding to a ferromagnetic phase \cite{marripour}. This difference arises from the entropy growth under a quasiperiodic drive: the symmetric interval, exhibits a significantly larger increase in entropy compared to the ordered ferromagnetic system, leading to faster destabilization of the prethermal TQC.

\section{Conclusion and Outlook}\label{Sec:Conclution}

In this study, we demonstrated the emergence and stability of a prethermal TQC phase in a spin-1/2 chain, subjected to a quasiperiodic drive. Through systematic analysis of the dynamical structure factor, time autocorrelation functions, and entanglement entropy via exact diagonalization, we identified a distinct QTTSB. This phase is characterized by robust, incommensurate frequency locking at $k\omega_d/2 \pm \Omega$ (for odd integers $k$) within an extended prethermal window.

A central finding is the pronounced suppression of heating in the high-frequency drive regime. In this limit, the half-chain EE exhibits slow, sublinear growth followed by a long-lived prethermal plateau. The corresponding prethermal lifetime $\tau_{\text{pre}}$ grows rapidly with increasing drive frequency, consistent with the theoretical expectation that resonant energy absorption is strongly suppressed. Analysis across accessible system sizes ($L = 2, 4, 6$) shows that the time autocorrelation function remains qualitatively unchanged, indicating that the observed prethermal stability is governed primarily by the high-frequency hierarchy rather than finite-size effects.

Importantly, the choice of disorder sampling interval plays a decisive role in the system's resistance to heating. A comparison between symmetric and asymmetric distributions reveals that an asymmetric distribution (with nonzero mean) places the system in a globally ordered ferromagnetic phase, which is highly resistant to energy absorption. In contrast, the standard symmetric interval leaves the system in a disordered phase, leading—particularly at lower frequencies—to more rapid heating and early breakdown of temporal order.

We further demonstrated that the TQC regime exhibits considerable robustness against moderate next-nearest-neighbor couplings and continuous control imperfections. However, due to the dense incommensurate frequency spectrum, the TQC phase possesses an inherent fragility compared to periodically driven discrete time crystals. Overall, our results establish disordered spin chains under quasiperiodic driving as a rich platform for realizing long-lived nonequilibrium temporal order, shedding light on the complex interplay between disorder and quasiperiodic drives.

Several open questions remain. First, beyond exact diagonalization, further investigation is required to understand the prethermal lifetime in larger systems and its precise scaling with driving frequency, as larger sizes demand greater computational resources. Determining the exact dependence of $\tau_{\text{pre}}$ on $\omega_d$ for quasiperiodic driving with irrational frequency ratios remains an open problem, as does the precise growth exponent of the entanglement entropy. Second, developing an effective analytical framework—possibly extending high-frequency expansion techniques to genuinely quasiperiodic drives, would help clarify the microscopic origin of heating suppression. Third, the role of disorder strength and the fate of the TQC regime in weakly disordered or clean systems remain open, along with the determination of the prethermalization plateau length and the critical points (in terms of nearest-neighbor coupling perturbations and rotation imperfections) that govern system stability. Finally, experimental realizations in programmable quantum simulators may provide direct access to quasiperiodic temporal order and its robustness.

In summary, our study provides a comprehensive foundation for understanding prethermal TQC dynamics, emphasizing how the interplay of high-frequency quasiperiodic driving and collective spin rigidity can effectively delay thermalization and preserve long-lived quasiperiodic order.
\section{Acknowledgment}

JA extends his appreciation to the Institute for Advanced Studies in Basic Sciences for their financial support through research Grant No. GIASBS12969.

\appendix
\section{Definition of quasiperiodic systems}\label{sec:app}

The Hamiltonian of a quasiperiodically driven system can be written as
\begin{equation}
	\hat{H}(t)=\sum_{\vec{n}} \hat{H}_{\vec{n}} \, 
	\exp\!\big(i\, \vec{n}\!\cdot\!\vec{\omega}\, t \big),
	\label{Eq:quasiperiod-Hamiltonian}
\end{equation}
where $\hat{H}_{\vec{n}}$ are time-independent operators, 
$\vec{\omega}=(\omega_1,\dots,\omega_d)$ is a $d$-dimensional vector of driving 
frequencies, and $\vec{n}=(n_1,\dots,n_d)$ is a vector of integers over which the sum 
runs\cite{r31,r38}.

When the frequencies $\omega_j$ are rationally related, the drive is effectively 
periodic and the dynamics repeat after a finite time set by the least common multiple 
of the individual driving periods. In contrast, if the $\omega_j$ are irrationally 
related, the Hamiltonian is genuinely quasiperiodic, leading to dynamics that never 
repeat exactly in time\cite{marripour}.

	A quasiperiodic Hamiltonian thus arises from a superposition of multiple oscillations with incommensurate frequencies. Unlike periodic systems, in quasiperiodic systems, the evolution explores a higher-dimensional torus in phase space, densely filling it without exact recurrences. This lack of strict periodicity is the defining feature of quasiperiodic driving and plays a central role in the emergence of nontrivial dynamical phases such as TQCs.
	In driven many-body systems, the notions of time-translation symmetry breaking (TTSB) 
	and its quasiperiodic analogue (QTTSB) provide a useful framework for characterizing 
	TC and TQC dynamical orders. For a periodically driven system with period $T$, TTSB occurs 
	when an observable develops a response with a period that is an integer multiple of $T$, 
	i.e., $\langle O(t+mT)\rangle = \langle O(t)\rangle$ for $m \ge 2$.
	
	In quasiperiodically driven systems, where no single fundamental period exists, time-translation symmetry breaking can instead be identified through the frequency content of the dynamics. Specifically, a quasiperiodic system demonstrates QTTSB behavior if there exists a local observable $\hat{O}$ whose expectation value satisfies the following Fourier expansion:
	\begin{equation}
		\langle \hat{O}(t)\rangle = \sum_{\vec n} O_{\vec n} \,
		e^{i (\vec n \cdot \vec{\tilde\omega}) t},
		\label{Eq:quasiperiod-def}
	\end{equation}
	where $\vec{n}$ is a $d$-dimensional vector of integers, and $\vec{\tilde{\omega}} = (\tilde{\omega}_1, \ldots, \tilde{\omega}_d)$ is a $d$-dimensional frequency vector with components defined as $\tilde{\omega}_j = \omega_j / p_j$. Here, $\omega_j$ are the fundamental driving frequencies, and $p_j$ are positive integers, with at least one $p_j > 1$ \cite{marripour,r38}. The coefficients $O_{\vec n}$ are the complex Fourier amplitudes corresponding to the quasiperiodic mode determined by $\vec{n}$. A non-vanishing amplitude $O_{\vec n}$ for modes involving fractional frequencies (arising from $p_j > 1$) serves as the direct dynamical signature of the QTTSB phase.



\begin{thebibliography}{99}

			
		\bibitem{r1}
		J. Dziarmaga,
		"Dynamics of a quantum phase transition and relaxation to a steady state'',
			\textit{Advances in Physics} {\bf 59}, 1063--1189 (2010).
			
		\bibitem{r2}
		A. Polkovnikov, K. Sengupta, A. Silva, and M. Vengalattore,
		"Colloquium: Nonequilibrium dynamics of closed interacting quantum systems'',
		\textit{Rev. Mod. Phys.} {\bf 83}, 863 (2011).
			
		\bibitem{r3}
		A. Dutta, G. Aeppli, B. K. Chakrabarti, U. Divakaran, T. F. Rosenbaum, and D. Sen,
		\textit{Quantum Phase Transitions in Transverse Field Spin Models: From Statistical Physics to Quantum Information}(Cambridge University Press, 2015).
			
		\bibitem{r4}
		A. Chandra, A. Das, and B. Chakrabarti,
		\textit{Quantum Quenching, Annealing and Computation}
		(Springer Berlin Heidelberg, 2010).
			
		\bibitem{r5}
		M. Bukov, L. D'Alessio, and A. Polkovnikov,
		"Universal high-frequency behavior of periodically driven systems: from dynamical stabilization to floquet engineering,''
		\textit{Advances in Physics} {\bf 64}, 139 (2015).
			
		\bibitem{r6}
		L. D'Alessio and A. Polkovnikov,
		"Many-body energy localization transition in systems,''
		\textit{Annals of Physics} {\bf 333}, 19 (2013).
			
		\bibitem{r7}
		L. D'Alessio, Y. Kafri, A. Polkovnikov, and M. Rigol,
		``From quantum chaos and eigenstate thermalization to statistical mechanics and thermodynamics,''
		\textit{Advances in Physics} {\bf 65}, 239 (2016).
			
		\bibitem{r8}
		S. Shevchenko, S. Ashhab, and F. Nori,
		``Landau-Zener-Stückelberg interferometry,''
		\textit{Physics Reports} {\bf 492}, 1--30 (2010).
			
		\bibitem{r9}
		T. Oka and S. Kitamura,
		``Floquet engineering of quantum materials,''
		\textit{Annual Review of Condensed Matter Physics} {\bf 10}, 387--408 (2019).
			
		\bibitem{r10}
		S. Blanes, F. Casas, J. Oteo, and J. Ros,
		``The magnus expansion and some of its applications,''
		\textit{Physics Reports} {\bf 470}, 151--238 (2009).
			
		\bibitem{r11}
		A. Eckardt,
		``Colloquium: Atomic quantum gases in periodically driven optical lattices,''
		\textit{Rev. Mod. Phys.} {\bf 89}, 011004 (2017).
			
		\bibitem{r12}
		A. Sen, D. Sen, and K. Sengupta,
		``Analytic approaches to periodically driven closed quantum systems: methods and applications,''
		\textit{Journal of Physics: Condensed Matter} {\bf 33}, 443003 (2021).
			
		\bibitem{r13}
		I. Bloch, J. Dalibard, and W. Zwerger,
		``Many-body physics with ultracold gases,''
		\textit{Rev. Mod. Phys.} {\bf 80}, 885 (2008).
			
		\bibitem{r14}
		L. Tarruell and L. Sanchez-Palencia,
		``Quantum simulation of the hubbard model with ultracold fermions in optical lattices,''
			\textit{Comptes Rendus. Physique} {\bf 19}, 365--393 (2018).
			
		\bibitem{r15}
		W. W. Ho, T. Mori, D. A. Abanin, and E. G. Dalla Torre,
		``Quantum and classical floquet prethermalization,''
		\textit{Annals of Physics} {\bf 454}, 169297 (2023).
			
		\bibitem{r16}
		T. Mori, T. N. Ikeda, E. Kaminishi, and M. Ueda,
		``Thermalization and prethermalization in isolated quantum systems: a theoretical overview,''
		\textit{Journal of Physics B: Atomic, Molecular and Optical Physics} {\bf 51}, 112001 (2018).
			
		\bibitem{r17}
		T. Banerjee and K. Sengupta,
		``Emergent symmetries in prethermal phases of periodically driven quantum systems,''
		\textit{Journal of Physics: Condensed Matter} {\bf 37}, 133002 (2025).
		\bibitem{px63-dtc9}
		Y.-B. Zhang, X.-C. Zhou, B.-Z. Wang, and X.-J. Liu,
		"Quantum Many-Body Dynamics for Fermionic $t\text{-}J$ Model Simulated with Atom Arrays",\textit{Phys. Rev. Lett}. \textbf{136}, 033402 (2026).
			
		\bibitem{r18} 
		F. Wilczek, "Quantum time crystals," \textit{Physical Review Letters} {\bf 109}, 160401 (2012).
		
		\bibitem{r19} 
		P. Bruno, Comment on “quantum time crystals”, \textit{Phys. Rev.
			Lett.},  \textbf{110}, 118901 (2013).
		\bibitem{r20} 
		H. Watanabe and M. Oshikawa, "Absence of quantum time crystals," \textit{Physical Review Letters} \textbf{114}, 251603 (2015).
		
		\bibitem{r21} 
		G. E. Volovik, On the broken time translation symmetry in
		macroscopic systems: Precessing states and off-diagonal longrange order,\textit{ JETP Lett}. \textbf{98}, 491 (2013).
		
		\bibitem{r22} 
		P. Bruno, "Impossibility of spontaneously rotating time-Crystals," \textit{Physical Review Letters} \textbf{111}, 070402 (2013).
		
		\bibitem{r23}
		J. Zhang et al., "Observation of a discrete time crystal,'' \textit{Nature} \textbf{543}, 217 (2017).
		
		\bibitem{r24}
		R. Yousefjani, A. Carollo, K. Sacha, S. Al-Kuwari, and A.
		Bayat, Non-Hermitian discrete time crystals, \textit{Phys. Rev. B} \textbf{111},165117 (2025).
		
		\bibitem{r25}
		R. Yousefjani, K. Sacha, and A. Bayat, "Discrete time crystal phase as a resource for quantum-enhanced sensing," \textit{Physical Review B} \textbf{111}, 125159 (2025).
		
		\bibitem{r26} 
		V. Khemani, A. Lazarides, R. Moessner, and S. L. Sondhi, "Phase structure of driven quantum systems," \textit{Physical Review Letters} \textbf{116}, 250401 (2016).
		
		\bibitem{r27} 
		D. V. Else, B. Bauer, and C. Nayak, Floquet time crystals," \textit{Physical Review Letters} \textbf{117}, 090402 (2016).
		
		\bibitem{marripour}
		D. Marripour and J. Abouie, "From time crystals to time quasicrystals: Exploring quasiperiodic phases in transverse field Ising chains," \textit{Physical Review B} \textbf{112}, 174307 (2025).
		
		\bibitem{Y. Huang}
		Y. Huang, T. Wang, H. Yin, M. Jiang, Z. Luo, and X. Peng, "Observation of continuous time crystals and quasi-crystals in spin gases," \textit{Nature Communications} \textbf{16}, 9375 (2025).
		\bibitem{Khemani2019}
		V. Khemani, R. Moessner, and S. L. Sondhi, "A brief history of time crystals," \textit{arXiv:1910.10745} (2019).
		\bibitem{r29}
		 D. V. Else, C. Monroe, C. Nayak, and N. Y. Yao, Discrete time crystals, \textit{Annual Review of Condensed Matter Physics} \textbf{11}, 467 (2020).
		 \bibitem{r30}
		 M. P. Zaletel, M. Lukin, C. Monroe, C. Nayak,
		 F. Wilczek, and N. Y. Yao, Colloquium: Quantum
		 and classical discrete time crystals,  \textit{Reviews of Modern Physics} \textbf{95}, 031001 (2023).
		

		 \bibitem{r31} Verdeny, J. Puig, and F. Mintert, Quasi-periodically driven quantum systems, \textit{Z. Naturforsch., A: Phys. Sci.} \textbf{71}, 897 (2016).
		 
		 \bibitem{r32} Nandy, A. Sen, and D. Sen, Aperiodically Driven Integrable Systems and Their Emergent Steady States, \textit{Phys. Rev. X} \textbf{7}, 031034 (2017).
		 
		 \bibitem{r33} Giergiel, A. Kuroś, and K. Sacha, Discrete time quasicrystals, \textit{Phys. Rev. B} \textbf{99}, 220303(R) (2019).
		 
		 \bibitem{r34}. Ray, S. Sinha, and D. Sen, Dynamics of quasiperiodically driven spin systems, \textit{Phys. Rev. E} \textbf{100}, 052129 (2019).
		 
		 \bibitem{r35} Zhao, F. Mintert, R. Moessner, and J. Knolle, Random Multipolar Driving: Tunably Slow Heating through Spectral Engineering, \textit{Phys. Rev. Lett.} \textbf{126}, 040601 (2021).
		 \bibitem{r36}
		 J Šuntajs and L. Vidmar, Ergodicity breaking transition in zero
		 dimensions, \textit{Phys. Rev. Lett.} \textbf{129}, 060602 (2022).
		 
		 \bibitem{r37}
		 S. Autti, V. B. Eltsov, and G. E. Volovik, "Observation of a time quasicrystal and its transition to a superfluid time crystal,'' \textit{Physical Review Letters} \textbf{120}, 215301 (2018).
		 
		 \bibitem{r38}
		 H. Zhao, F. Mintert, and J. Knolle, "Floquet time spirals and stable discrete-time quasicrystals in quasiperiodically driven quantum many-body systems," \textit{Physical Review B} \textbf{100}, 13 (2019).
		 
		 \bibitem{r39}
		 X. Wu, "Quasi time crystal," \textit{arXiv preprint arXiv:2203.01468} (2022).
		 
		 \bibitem{r40}
		 G. He et al., "Experimental realization of discrete time quasi-crystals," \textit{arXiv preprint arXiv:2403.17842} (2024).
		 
		 \bibitem{r41}. Mukherjee, A. Sen, D. Sen, and K. Sengupta, Restoring coherence via aperiodic drives in a many-body quantum system, \textit{Phys. Rev. B} \textbf{102}, 014301 (2020).
		 
		 \bibitem{r42}
		 H. Schmid, et al., "Self-similar phase diagram of the Fibonacci-driven quantum Ising model," arXiv:2410.18219 (2024).
		 
		 \bibitem{r43}
		 Y. Peng and G. Refael, "Time-quasiperiodic topological superconductors with Majorana multiplexing," \textit{Physical Review B} \textbf{98}, 220509 (2018).
		 
		 \bibitem{r44}
		 X. Luo, Y. Zhou, Z. Xu, and W. Jiang, "Discrete time quasi-crystal in Rydberg atomic chain," arXiv:2505.09117 (2025).
		 
  		\bibitem{das2010}
  		A. Das,
  		``Exotic freezing of response in a quantum many-body system,''
  		\textit{Phys. Rev. B} {\bf 82}, 172402 (2010).
  		
  		\bibitem{bhattacharyya2012}
  		S. Bhattacharyya, A. Das, and S. Dasgupta,
  		``Transverse Ising chain under periodic instantaneous quenches: Dynamical many-body freezing and emergence of slow solitary oscillations,''
  		\textit{Phys. Rev. B} {\bf 86}, 054410 (2012).
  		
  		\bibitem{hegde2014}
  		S. S. Hegde, H. Katiyar, T. S. Mahesh, and A. Das,
  		``Freezing a quantum magnet by repeated quantum interference: An experimental realization,''
  		\textit{Phys. Rev. B} {\bf 90}, 174407 (2014).
  		
  		
  		\bibitem{mondal2012}
  		S. Mondal, D. Pekker, and K. Sengupta,
  		``Dynamics-induced freezing of strongly correlated ultracold bosons,''
  		\textit{EPL (Europhysics Letters)} {\bf 100}, 60007 (2012).
  		
  		\bibitem{guo2025}
  		H. Guo, R. Mukherjee, and D. Chowdhury,
  		``Dynamical freezing in exactly solvable models of driven chaotic quantum dots,''
  		\textit{Phys. Rev. Lett.} {\bf 134}, 184302 (2025).
  		
  		\bibitem{divakaran2014}
  		U. Divakaran and K. Sengupta,
  		``Dynamic freezing and defect suppression in the tilted one-dimensional Bose-Hubbard model,''
  		\textit{Phys. Rev. B} {\bf 90}, 184303 (2014).
  		
  		\bibitem{camilo2020}
  		G. Camilo and D. Teixeira,
  		``Complexity and Floquet dynamics: Nonequilibrium Ising phase transitions,''
  		\textit{Phys. Rev. B} {\bf 102}, 174304 (2020).
		
		\bibitem{mukherjee2024}
		B. Mukherjee, R. Melendrez, M. Szyniszewski, H. J. Changlani, and A. Pal,
		``Emergent strong zero mode through local Floquet engineering,''
		\textit{Phys. Rev. B} {\bf 109}, 064303 (2024).
		
		\bibitem{koch2023}
		J. Koch, G. R. Hunanyan, T. Ockenfels, E. Rico, E. Solano, and M. Weitz,
		``Quantum Rabi dynamics of trapped atoms far in the deep strong coupling regime,''
		\textit{Nat. Commun.} {\bf 14}, 954 (2023).
		
		\bibitem{haldar2021}
		A. Haldar, D. Sen, R. Moessner, and A. Das,
		``Dynamical freezing and scar points in strongly driven : Resonance vs emergent conservation laws,''
		\textit{Phys. Rev. X} {\bf 11}, 021008 (2021).
		
		\bibitem{banerjee2023a}
		T. Banerjee and K. Sengupta,
		``Emergent conservation in the Floquet dynamics of integrable non-Hermitian models,''
		\textit{Phys. Rev. B} {\bf 107}, 155117 (2023).
		
		\bibitem{banerjee2024}
		T. Banerjee and K. Sengupta,
		``Entanglement transitions in a periodically driven non-Hermitian Ising chain,''
		\textit{Phys. Rev. B} {\bf 109}, 094306 (2024).
		
		\bibitem{turkeshi2023}
		X. Turkeshi and M. Schir\'o,
		``Entanglement and correlation spreading in non-Hermitian spin chains,''
		\textit{Phys. Rev. B} {\bf 107}, L020403 (2023).
		\bibitem{gangopadhay2025}
		N. Gangopadhay and S. Choudhury,
		``Counterdiabatic route to entanglement steering and dynamical freezing in the Floquet Lipkin-Meshkov-Glick model,''
		\textit{Phys. Rev. Lett.} {\bf 135}, 020407 (2025).
		
	

		
		\bibitem{pai2019}
		S. Pai and M. Pretko,
		``Dynamical scar states in driven fracton systems,''
		\textit{Phys. Rev. Lett.} {\bf 123}, 136401 (2019).
		
		\bibitem{mukherjee2020a}
		B. Mukherjee, S. Nandy, A. Sen, D. Sen, and K. Sengupta,
		``Collapse and revival of quantum many-body scars via Floquet engineering,''
		\textit{Phys. Rev. B} {\bf 101}, 245107 (2020).
		
		\bibitem{mizuta2020}
		K. Mizuta, K. Takasan, and N. Kawakami,
		``Exact Floquet quantum many-body scars under Rydberg blockade,''
		\textit{Phys. Rev. Res.} {\bf 2}, 033284 (2020).
		\bibitem{H. Yarloo} 
		H. Yarloo, A. Emami Kopaei, and A. Langari, "Homogeneous Floquet time crystal from weak ergodicity breaking," \textit{Physical Review B} \textbf{102}, 224309 (2020).
		
		\bibitem{sugiura2021}
		S. Sugiura, T. Kuwahara, and K. Saito,
		``Many-body scar state intrinsic to periodically driven system,''
		\textit{Phys. Rev. Res.} {\bf 3}, L012010 (2021).
		
		\bibitem{mukherjee2020b}
		B. Mukherjee, A. Sen, D. Sen, and K. Sengupta,
		``Dynamics of the vacuum state in a periodically driven Rydberg chain,''
		\textit{Phys. Rev. B} {\bf 102}, 075123 (2020).
		
		\bibitem{maskara2021}
		N. Maskara, A. A. Michailidis, W. W. Ho, D. Bluvstein, S. Choi, M. D. Lukin, and M. Serbyn,
		``Discrete time-crystalline order enabled by quantum many-body scars: Entanglement steering via periodic driving,''
		\textit{Phys. Rev. Lett.} {\bf 127}, 090602 (2021).
		
		\bibitem{hudomal2022}
		A. Hudomal, J.-Y. Desaules, B. Mukherjee, G.-X. Su, J. C. Halimeh, and Z. Papi\'c,
		``Driving quantum many-body scars in the PXP model,''
		\textit{Phys. Rev. B} {\bf 106}, 104302 (2022).
		
		\bibitem{huang2022}
		B. Huang, T.-H. Leung, D. M. Stamper-Kurn, and W. V. Liu,
		``Discrete time crystals enforced by Floquet-Bloch scars,''
		\textit{Phys. Rev. Lett.} {\bf 129}, 133001 (2022).
		
		
	
		
       \bibitem{B. Bauer} 
        D. V. Else, B. Bauer, and C. Nayak, "Prethermal phases of matter protected by time-translation symmetry," \textit{Physical Review X} \textbf{7}, 011026 (2017).
		\bibitem{K. Mallayya} 
		K. Mallayya, M. Rigol, and W. De Roeck, "Prethermalization and thermalization in isolated quantum systems," \textit{Physical Review X} \textbf{9}, 021027 (2019).
	
		\bibitem{D. V. Else} D. V. Else, W. W. Ho, and P. T. Dumitrescu, Long-lived Interacting Phases of Matter Protected by Multiple Time-Translation Symmetries in Quasiperiodically Driven Systems, \textit{Phys. Rev. X} \textbf{10}, 021032 (2020).
		\bibitem{G. He} 
		G. He et al., "Quasi-Floquet prethermalization in a disordered dipolar spin ensemble in diamond," \textit{Physical Review Letters} \textbf{131}, 130401 (2023).
		\bibitem{S. A. Weidinger} 
		S. A. Weidinger and M. Knap, "Floquet prethermalization and regimes of heating in a periodically driven, interacting quantum system," \textit{Scientific Reports} \textbf{7}, 45382 (2017).
		\bibitem{E. Canovi} 
		E. Canovi, M. Kollar, and M. Eckstein, "Stroboscopic prethermalization in weakly interacting periodically driven systems," \textit{Physical Review E} \textbf{93}, 012130 (2016).
		\bibitem{M. Bukov} 
		M. Bukov, S. Gopalakrishnan, M. Knap, and E. Demler, "Prethermal Floquet steady states and instabilities in the periodically driven, weakly interacting Bose-Hubbard model," \textit{Physical Review Letters} \textbf{115}, 205301 (2015).
		 
		 \bibitem{Kyprianidis}
		 A. Kyprianidis, F. Machado, W. Morong, P. Becker, K. S. Collins, D. V. Else, ..., and C. Monroe, "Observation of a prethermal discrete time crystal," \textit{Science} \textbf{372}, 1192--1196 (2021).
		 \bibitem{Das Sarma}
		 D. Vu and S. Das Sarma, "Dissipative prethermal discrete time crystal," \textit{Physical Review Letters} \textbf{130}, 130401 (2023)
		\bibitem{Zeng}
		T. S. Zeng and D. N. Sheng, "Prethermal time crystals in a one-dimensional periodically driven Floquet system," \textit{Physical Review B} \textbf{96}, 094202 (2017).
		\bibitem{Stasiuk}
		 A. Stasiuk and P. Cappellaro, "Observation of a prethermal U(1) discrete time crystal," \textit{Physical Review X} \textbf{13}, 041016 (2023).
		 \bibitem{P. T. Dumitrescu}
		 P. T. Dumitrescu, R. Vasseur, and A. C. Potter, "Logarithmically slow relaxation in quasiperiodically driven random spin chains," \textit{Physical Review Letters} \textbf{120}, 070602 (2018).
		 
		 
		 \bibitem{Nandkishore}
		 R. Nandkishore, S. Gopalakrishnan, and D. A. Huse, "Spectral features of a many-body-localized system weakly coupled to a bath," \textit{Physical Review B} \textbf{90}, 064203 (2014).
		 \bibitem{Lazarides} 
		 A. Lazarides, A. Das, and R. Moessner, "Fate of many-body localization under periodic driving," \textit{Physical Review Letters} \textbf{115}, 030402 (2015).
		 \bibitem{Kjall}
		 J. A. Kjäll, J. H. Bardarson, and F. Pollmann, "Many-body localization in a disordered quantum Ising chain," \textit{Physical Review Letters} \textbf{113}, 107204 (2014).
		 \bibitem{huse}
		 Arijeet Pal and David A. Huse, “Many-body localization
		 phase transition,” \textit{Phys. Rev. B} \textbf{82}, 174411 (2010).
		 \bibitem{Bordia}
		 P. Bordia, H. Lüschen, U. Schneider, M. Knap, and I. Bloch, "Periodically driving a many-body localized quantum system," \textit{Nature Physics} \textbf{13}, 460--464 (2017).
		 
		 \bibitem{Oganesyan}
		 V. Oganesyan and D. A. Huse, "Localization of interacting fermions at high temperature," \textit{Physical Review B} \textbf{75}, 155111 (2007).
		  \bibitem{Keyserlingk}
		 C. W. von Keyserlingk and S. L. Sondhi, "Phase structure of one-dimensional interacting Floquet systems. II. Symmetry-broken phases," \textit{Physical Review B} \textbf{93}, 245146 (2016).
		  \bibitem{Johri}
		 S. Johri, R. Nandkishore, and R. N. Bhatt, "Many-body localization in imperfectly isolated quantum systems," \textit{Physical Review Letters} \textbf{114}, 117401 (2015).
		  \bibitem{Huse}
		 D. A. Huse, R. Nandkishore, V. Oganesyan, A. Pal, and S. L. Sondhi, "Localization-protected quantum order," \textit{Physical Review B} \textbf{88}, 014206 (2013).
		 \bibitem{P. Ponte}
		 P. Ponte, Z. Papić, F. Huveneers, and D. A. Abanin, "Many-body localization in periodically driven systems," \textit{Phys. Rev. Lett.} \textbf{114}, 140401 (2015).
		 
		 
		 \bibitem{Igloi2007}
		 F. Iglói, R. Juhász, and Z. Zimborás, 
		 ``Entanglement entropy of aperiodic quantum spin chains,'' 
		 Europhys. Lett. \textbf{79}, 37001 (2007).
		 
		 
	
		 
		 
		 \bibitem{Langlett}
		 C. M. Langlett and S. Xu, "Hilbert space fragmentation and exact scars of generalized Fredkin spin chains," \textit{Physical Review B} \textbf{103}, L220304 (2021).
		 \bibitem{Moudgalya}
		 S. Moudgalya and O. I. Motrunich, "Hilbert space fragmentation and commutant algebras," \textit{Physical Review X} \textbf{12}, 011050 (2022).
		 \bibitem{Moudgalya1}
		 S. Moudgalya, B. A. Bernevig, and N. Regnault, "Quantum many-body scars and Hilbert space fragmentation: A review of exact results," \textit{Reports on Progress in Physics} \textbf{85}, 086501 (2022).
		 
		 
		 \bibitem{Y. Baum}
		 Y. Baum, E. van Nieuwenburg, and G. Refael, "From dynamical localization to bunching in interacting floquet systems," \textit{SciPost Physics} \textbf{5}, 017 (2018).
		
		 \bibitem{T. Nag}
		 T. Nag, S. Roy, A. Dutta, and D. Sen, "Dynamical localization in a chain of hard core bosons under periodic driving," \textit{Physical Review B} \textbf{89}, 165425 (2014).
	
		 
		 \bibitem{Luitz}
		 D. J. Luitz, Y. Bar Lev, and A. Lazarides, "Absence of dynamical localization in interacting driven systems," \textit{SciPost Physics} \textbf{3}, 029 (2017).
		 
		 \bibitem{Agarwala}
		 A. Agarwala and D. Sen, "Effects of interactions on periodically driven dynamically localized systems," \textit{Physical Review B} \textbf{95}, 014305 (2017).
	
		 
		
		 \bibitem{L. Tamang}
		 L. Tamang, T. Nag, and T. Biswas, "Floquet engineering of low-energy dispersions and dynamical localization in a periodically kicked three-band system," \textit{Physical Review B} \textbf{104}, 174308 (2021).
		 
		 \bibitem{M. Fava}
		 M. Fava, R. Fazio, and A. Russomanno, "Many-body dynamical localization in the kicked bose-hubbard chain," \textit{Physical Review B} \textbf{101}, 064302 (2020).
		 
		 \bibitem{Keser}
		 A. C. Keser, S. Ganeshan, G. Refael, and V. Galitski, "Dynamical many-body localization in an integrable model," \textit{Physical Review B} \textbf{94}, 085120 (2016).
		 
		 \bibitem{Martinez}
		 D. F. Martinez, "High-order harmonic generation and dynamic localization in a driven two-level system, a nonperturbative solution using the floquet-green formalism," \textit{Journal of Physics A: Mathematical and General} \textbf{38}, 9979--10005 (2005).
		  	\bibitem{Lazarides2014}
		 A. Lazarides, A. Das, and R. Moessner, "Equilibrium states of generic quantum systems subject to periodic driving," \textit{Physical Review E} {\bf 90}, 012110 (2014).
		 
		 \bibitem{DAlessio2014}
		 L. D'Alessio and M. Rigol, "Long-time behavior of isolated periodically driven interacting lattice systems," \textit{Physical Review X} {\bf 4}, 041048 (2014).
		  \bibitem{Martin2021}
		 I. Martin and B. Z. Spivak, "Quasiperiodic Floquet states," \textit{Physical Review B} {\bf 103}, 075118 (2021).
		 \bibitem{V. Tiwari}
		 V. Tiwari, D. S. Bhakuni, and A. Sharma, "Periodically and aperiodically Thue-Morse driven long-range systems: From dynamical localization to slow dynamics," \textit{Physical Review B} \textbf{111}, 205109 (2025).
		 \bibitem{Zhao2021}
		 H. Zhao, F. Mintert, R. Moessner, and J. Knolle, ``: Tunably slow heating through spectral engineering,'' \textit{Physical Review Letters} \textbf{126}, 040601 (2021).
		 \bibitem{M. Serbyn}
		 M. Serbyn, Z. Papić, and D. A. Abanin, "Universal slow growth of entanglement in interacting strongly disordered systems," \textit{Physical Review Letters} \textbf{110}, 260601 (2013).
		 \bibitem{X. Wen}
		 X. Wen, Y. Gu, A. Vishwanath, and R. Fan, "Periodically, quasi-periodically, and randomly driven conformal field theories (II): Furstenberg's theorem and exceptions to heating phases," \textit{SciPost Physics} \textbf{13}, 082 (2022).
		 
		\bibitem{Van Hove}
		L. Van Hove, "Correlations in space and time and Born approximation scattering in systems of interacting particles," \textit{Phys. Rev.} \textbf{95}, 249--262 (1954).
		\bibitem{Pragna Das}
		Pragna Das, Devendra Singh Bhakuni, Lea F. Santos, and Auditya Sharma,"Periodically and quasiperiodically driven anisotropic Dicke model," \textit{Phys. Rev. A} \textbf{108}, 063716,(2023).
		\bibitem{Takashi Mori}
		Takashi Mori, Hongzheng Zhao,eta al,"Rigorous Bounds on the Heating Rate in Thue-Morse Quasiperiodically and Randomly Driven Quantum Many-Body Systems," \textit{Phys. Rev. Lett} \textbf{127}, 050602 (2021).
		\bibitem{Calabrese2004}
		P. Calabrese and J. Cardy, 
		``Entanglement entropy and quantum field theory,'' 
		J. Stat. Mech. \textbf{2004}, P06002 (2004).
		
		\bibitem{Latorre2004}
		J. I. Latorre, E. Rico, and G. Vidal, 
		``Ground state entanglement in quantum spin chains,'' 
		Quantum Inf. Comput. \textbf{4}, 48 (2004).
		\bibitem{Page1993}
		D. N. Page, "Average entropy of a subsystem," \textit{Phys. Rev. Lett.} \textbf{71}, 1291--1294 (1993).
		\bibitem{Abanin2015}
		D. A. Abanin, W. De Roeck, and F. Huveneers, "Exponentially slow heating in periodically driven many-body systems," \textit{Physical Review Letters} {\bf 115}, 256803 (2015).
		
		
		
		
		
	\end{thebibliography}
\end{document}